\documentclass[natbib]{cambridge6A}
\usepackage{graphicx}
\usepackage{times}
\def\la{\mathrel{\hbox{\rlap{\hbox{\lower4pt\hbox{$\sim$}}}{\raise2pt\hbox{$<$}}}}}
\def\ga{\mathrel{\hbox{\rlap{\hbox{\lower4pt\hbox{$\sim$}}}{\raise2pt\hbox{$>$}}}}}
\def\ls{\mathrel{\hbox{\rlap{\hbox{\lower4pt\hbox{$\sim$}}}\hbox{$<$}}}}
\def\gs{\mathrel{\hbox{\rlap{\hbox{\lower4pt\hbox{$\sim$}}}\hbox{$>$}}}}

\begin{document}

\title{Observational Characteristics of Accretion onto Black Holes}
\author{C. Done\\[3\baselineskip]
Department of Physics, University of Durham, South Road, Durham, DH1
3LE: \\[1\baselineskip]
chris.done@durham.ac.uk} 
\maketitle

\chapter{Observational Characteristics of Accretion onto Black Holes}

\section{Abstract}

These notes resulted from a series of lectures at the IAC winter
school. They are designed to help students, especially those just
starting in subject, to get hold of the fundamental tools used to
study accretion powered sources. As such, the references give a place
to start reading, rather than representing a complete survey of
work done in the field.

I outline Compton scattering and blackbody radiation as the two
predominant radiation mechanisms for accreting black holes, producing
the hard X-ray tail and disc spectral components, respectively. The
interaction of this radiation with matter can result in photo-electric
absorption and/or reflection.  While the basic processes can be found
in any textbook, here I focus on
how these can be used as a toolkit to interpret the spectra and
variability of black hole binaries (hereafter BHB) and Active Galactic
Nuclei (AGN). I also discuss how to use these to physically interpret
real data using the publicly available {\sc xspec} spectral fitting
package  (Arnaud et al 1996), and how this has led to current models (and controversies) of
the accretion flow in both BHB and AGN.

\section{Fundamentals of accretion flows: observation and theory}

\subsection{Plotting Spectra}
\label{s:plot}

Spectra can often be (roughly) represented as a power law. This can be
written as a differential photon number density (photons per second
per square cm per energy band) as $N(E)=N_0 E^{-\Gamma}$ where
$\Gamma$ is photon index. The energy flux is then simply $F(E)=EN(E) =
N_0 E^{-(\Gamma-1)}= N_0 E^{-\alpha}$ where $\alpha=\Gamma-1$ is
energy index. 

Power law spectra are broad band, i.e. the emission spans many decades
in energy. Thus in general we plot logarithmically, in $\log E$, with
$d\log E$ rather than $dE$ as the constant.  The number of photons per
bin is $N(E)dE =N(E)E dE/E=EN(E)d\log E=F(E)d\log E$. Thus, somewhat
counter-intuitively, plotting $F(E)$ on a logarithmic energy scale 
shows the {\em number} of photons rather than flux.

\begin{figure*}[!h]
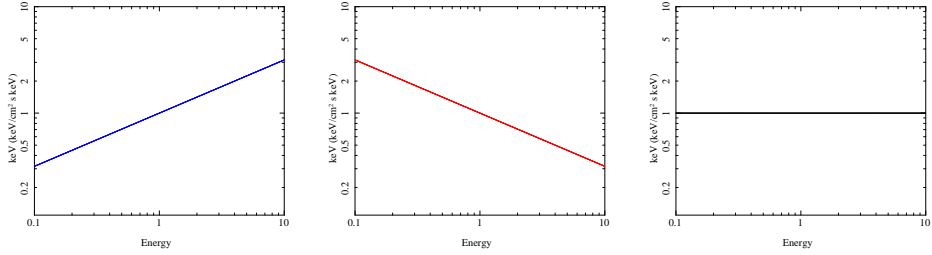

\begin{tabular}{ccc}
\includegraphics[width=0.3\textwidth]{hard.ps}
&
\includegraphics[width=0.3\textwidth]{soft.ps}
&
\includegraphics[width=0.3\textwidth]{flat.ps}

\end{tabular}
\caption{The $\nu f(\nu)$ spectra of a power law with photon index
$\Gamma$ of 1.5, 2.5 and 2.0. This representation makes it clear that
the power output peaks at high energies for hard spectra ($\Gamma<2$)
while it peaks at low energies for soft spectra ($\Gamma>2$). }
\label{f:pl}
\end{figure*}

Similarly, energy per bin is $F(E) dE =F(E)E dE/E = EF(E) d\log E$.
Thus to see the energy at which the source luminosity peaks 
on a logarithmic frequency scale means we have to plot 
$\log E F(E)$ ($=\nu F(\nu)$) versus $\log E$. 
In these units, hard spectra have
$\Gamma<2$ so peak at high energies.  Soft spectra have $\Gamma > 2$
and peak at low energies.  Flat spectra with $\Gamma=2$ means equal
power per decade. These are illustrated in Fig.~\ref{f:pl}. 

One other very common way for spectra to be plotted is in counts per
second per cm$^2$, $C(E)$. This is {\em not} the same as $N(E)$, the
counts refer to number of photons {\em detected} not emitted. Such
spectra show more about the detector than the intrinsic spectrum as
these are convolved with the detector response giving $C(E)=\int
N(E_0) R(E,E_0) dE_0$ where $R(E,E_0)$ is the detector response
i.e. the probability that a photon of input energy $E_0$ is detected
at energy $E$. Instrument responses for X-rays are generally complex
so the spectra are generally analysed in counts space, by convolving
models of the intrinsic spectrum through the detector response and
minimizing the difference between the predicted and detected counts to
derive the best fit model parameters. While such 'counts spectra'
constitute the basic observational data, they do not give a great deal
of physical insight. Deconvolving these data using a model to plot
them in $\nu f(\nu)$ space is strongly recommended i.e. in {\sc xspec} using 
the command {\sc iplot eeuf}.

\subsection{Plotting variability}
\label{s:pow}

Plotting variability is analogous to plotting spectra. A light-curve,
$I(t)$, spanning time $T$, with points every $\Delta t$ can be
decomposed into a sum of sinusoids:
$$I(t)=I_0+\Sigma_{i=1}^N A_i\sin (2\pi \nu_i t + \phi_i)$$
where $I_0$ is the average flux over that timescale, $\nu_i=i/T$ with
$i=1,2...N$ and $N=T/(2\Delta t)$.  This is more useful when
normalized to the average, giving the fractional change in intensity
$I(t)/I_0= 1+ \Sigma (A_i/I_0)\sin (2\pi \nu_i t + \phi_i)$. The power
spectrum $P(\nu )$ is $(A_i/I_0)^2$ versus $\nu_i$, and the
integral $\int P(\nu )d\nu = (\sigma/I)^2$ i.e. the squared 
total r.m.s. variability
of the lightcurve.

Similar to the energy spectra, the power spectrum is generally broad
band so is plotted logarithmically. Then the total variability power
in a bin is $P(\nu ) d\nu = P(\nu ) \nu d\nu/\nu  = \nu P(\nu )
d\log\nu$. Thus, similarly
to spectra, a peak in $\nu P(\nu )$ versus $\log\nu$ shows the frequency at
which the variability power peaks, so this is the more physical way to
plot power spectra. 

\begin{figure*}
\begin{tabular}{c}
\includegraphics[width=0.8\textwidth,angle=-90]{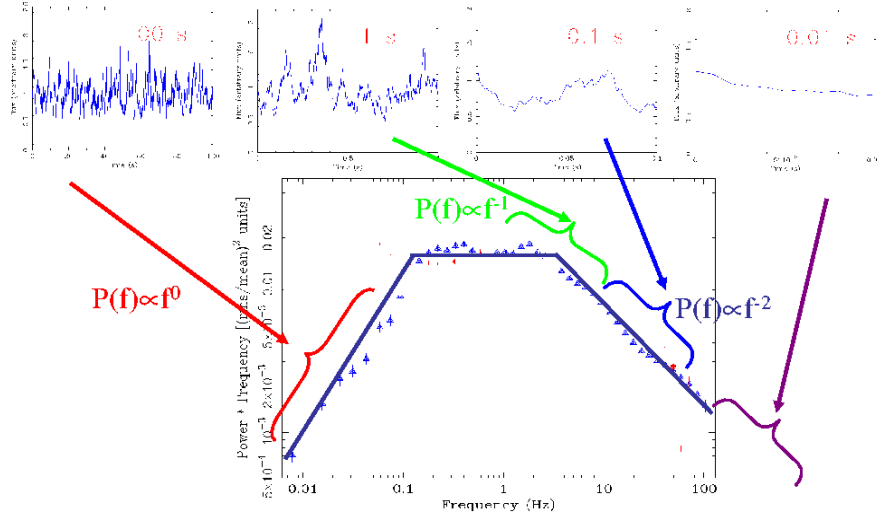}
\end{tabular}
\caption{The lower panel shows the power spectrum of a lightcurve 
from a low/hard state of Cyg X-1. 
The upper panels show segements of the lightcurve spanning timescales of
100s, 1s, 0.1s and 0.01s, respectively. These are
each normalized to their mean, so it is easy to see that the
fractional variability is largest in the 1s light-curve, and that this 
is what the power spectrum measures. Figure courtesy of P. Uttley.
(Uttley 2007)}
\label{f:pow}
\end{figure*}

The data often show power-law power spectra, with $P(\nu ) \propto \nu^0$
on long timescales breaking to $\nu^{-1}$ at a low frequency break $\nu_b$
and then breaking again to $\nu^{-2}$ at a high frequency break
$\nu_h$. This is termed band limited noise or flat-top noise, where the
'flat-top' has $P(\nu ) \propto \nu^{-1}$ as this has equal variability
power per decade.

Fig.\ref{f:pow} illustrates this for data from a low/hard state (see
section \ref{s:states}) in Cyg X-1. The
light-curves are shown over different timescales, $T$.
These are all normalized to their mean, so it is
easy to see from the light-curves that the fractional variability is
very low on timescales shorter than 0.01~s. At longer timescales the
fractional variability increases, then remains constant for 1-10~s and
then drops again.

\subsection{Spectra and Variability of the Shakura-Sunyaev disc}
\label{s:ss}

The underlying physics of a Shakura-Sunyaev accretion disc can be
illustrated in a very simple derivation just conserving energy (rather
than the proper derivation which conserves energy and angular
momentum).

A mass accretion rate $\dot{M}$ spiraling inwards from $R$ to $R-dR$
liberates potential energy at a rate
$dE/dt=L_{pot}=(GM\dot{M}/R^2)\times dR$. The virial theorem says that only
half of this can be radiated, so $dL_{rad}=GM\dot{M} dR/(2R^2) $. If this
thermalises to a blackbody then $dL=dA \sigma_{SB} T^4$ where
$\sigma_{SB}$ is the Stephan-Boltzman constant and area of the annulus 
$dA=2 \times 2
\pi R \times dR$ (where the factor 2 comes from the fact that there is a top
and bottom face of the ring).  Then the luminosity from the annulus
$dL_{rad}=GM\dot{M}dR/(2R^2) = 4 \pi
R \times dR \sigma_{SB} T^4$ or $\sigma_{SB}T^4(R)= GM\dot{M}/8 \pi R^3 $.
This is only out by a factor $3(1-(R_{in}/R)^{1/2})$ which comes from a
full analysis including angular momentum (see H. Spruit in these
proceedings).

Thus the spectrum from a disc is a sum of blackbody components, with
increasing temperature and luminosity emitted from a decreasing area
as the radius decreases.  The peak luminosity and temperature then
comes from $R_{in}$ (modulo the corrections for the inner boundary
condition). Using the very approximate treatment above, the total 
total luminosity of the disc $ L_{disc} = GM\dot{M}/(2R_{in})$ so 
substituting for $GM\dot{M}$ gives 
$\sigma_{SB} T^4(R)= R_{in} L_{disc} /4 \pi R^3 \propto
L_{disc}\times (R_{in}/R) \times 1/(4 \pi R^2)$. So $\sigma_{SB}
T^4_{max}=\sigma_{SB} T(R_{in})= L_{disc}/(4 \pi R_{in}^2)$, giving an
observational constraint on $R_{in}$ if we can measure $T_{max}$ and
$L_{disc}$. This is important as $R_{in}$ is set by General Relativity
at the last stable orbit around the black hole, which is itself
dependent on spin.  Angular momentum, $J$, is typically a mass,
times a velocity, times a size scale. The smallest size scale for a
black hole is that of the event horizon, which is always larger than
$R_g=GM/c^2$, and the fastest velocity is the speed of light. Thus
$|J| < M c GM/c^2$, or spin-per-unit mass, $|J|/M = a_* GM/c$,
where $a_*\le 1$. The last stable orbit is at $6~R_g$ for a zero spin
($a_*=0$: Schwarzschild) black hole, decreasing to $1~R_g$ for a
maximally rotating Kerr black hole for the disc co-rotating with the
black hole spin ($a_*=1$), or $9~R_g$ for a counter-rotating disc
($a_*=-1$).  Thus to convert the observed emission area to spin we
need to know the mass of the black hole so we can put the observed
inner radius into gravitational radii $r_{in}=R_{in}/R_g$, giving us a way to
observationally measure black hole spin.

In {\sc xspec}, a commonly used model for the disc is {\sc
diskbb}. This assumes that $T^4\propto r^{-3}$ i.e. has no inner
boundary condition. It is adequate to fit the high energy part of the
disc spectrum i.e. the peak and Wien tail, but the derived
normalization needs to be corrected for the lack of boundary
condition. It also assumes that each radius emits as a true
blackbody, which is only true if the disk is effectively optically
thick to absorption at all frequencies. Free-free (continuum)
absorption drops as a function of frequency, so the highest energy
photons from each radii are unlikely to thermalize. This forms instead
a modified (or diluted) blackbody, with effective temperature which is
a factor $f_{\rm col}$ (termed a colour temperature correction) higher
than for complete thermalization. The full disk spectrum is then a sum
of these modified blackbodies, but this can likewise be approximately
described by a single colour temperature correction to a 'sum of
blackbodies' disk spectrum (Shimura \& Takahara 1995), giving rise to
a further correction to the {\sc diskbb} normalization. The final
factor is that the emission from each radius is smeared out by the
combination of special and general relativistic effects which arise
from the rapid rotation of the emitting material in a strong
gravitational field (Cunningham 1975, see section \ref{s:rel}). Again,
these corrections can be applied to the {\sc diskbb} model
(e.g. Kubota et al 2001; Gierlinski \& Done 2004a), but are only easily
available as tabulated values for spin $0$ and $0.998$ in Zhang et al
(1997).  Hence a better approach is to use {\sc kerrbb}, which
incorporates the stress-free boundary condition and relativistic
smearing for any spin (Li et al 2005) for a given colour temperature
correction factor. An even better approach is to use {\sc bhspec},
which calculates the intrinsic spectrum from each radius using full
radiative transfer through the disc atmosphere, including partially
ionized metal opacities, rather than assuming a colour temperature
corrected blackbody form (Davis et al 2005). This imprints atomic
features onto the emission from each radius, distorting the spectrum
from the smooth continuum as produced by {\sc kerrbb} (Done \& Davis
2008, Kubota et al 2010).

To zeroth order, 
the emitted spectrum does not require any assumptions about the nature
of the viscosity, parameterized by $\alpha$ by Shakura \& Sunyaev
(1973). However, variability is dependent on this, as variability in
the emitted spectrum requires that the mass accretion rate through the
disc changes. Material can only fall in if its angular momentum is
transported outwards via 'viscous' stresses, now known to be 
due to the magneto-rotational instability (MRI,
see J. Hawley lecture notes in this proceedings).  The viscous timescale,
$t_{visc}\approx \alpha^{-1} (H/R)^{-2} t_{dyn}$ where $H$ is the vertical
scale height of the disc and $t_{dyn}= 2\pi R_g(r^{3/2}+a_*)/c$ is the
dynamical (orbital) timescale which is $\sim 5$~ms for a Schwarzschild 
black hole of $10M_\odot$. 

Modeling the observed variability of the disc gives an estimate for
$\alpha=0.1$ (King, Pringle \& Livio 2007), though current simulations of 
the MRI gives stresses which are an order of magnitude lower than this
(see J. Hawley lecture notes in this proceedings).

The disc models give a
geometrically thin solution $H/R \sim 0.01$, so the very fastest
variability from changes in mass accretion rate at the innermost edge
are 100,000 times longer than the dynamical timescale. Thus 
accretion discs in black hole binaries (hereafter BHB) should only
vary on timescales longer than a few hundred seconds.

\subsection{Observed Spectra and Power Spectra of Black Hole Binaries}
\label{s:states}

\subsubsection{Disc dominated states}
\label{s:lt}

\begin{figure*}[!h]
\begin{tabular}{c}
\includegraphics[width=0.9\textwidth, bb=32 530 554 757,clip=true]{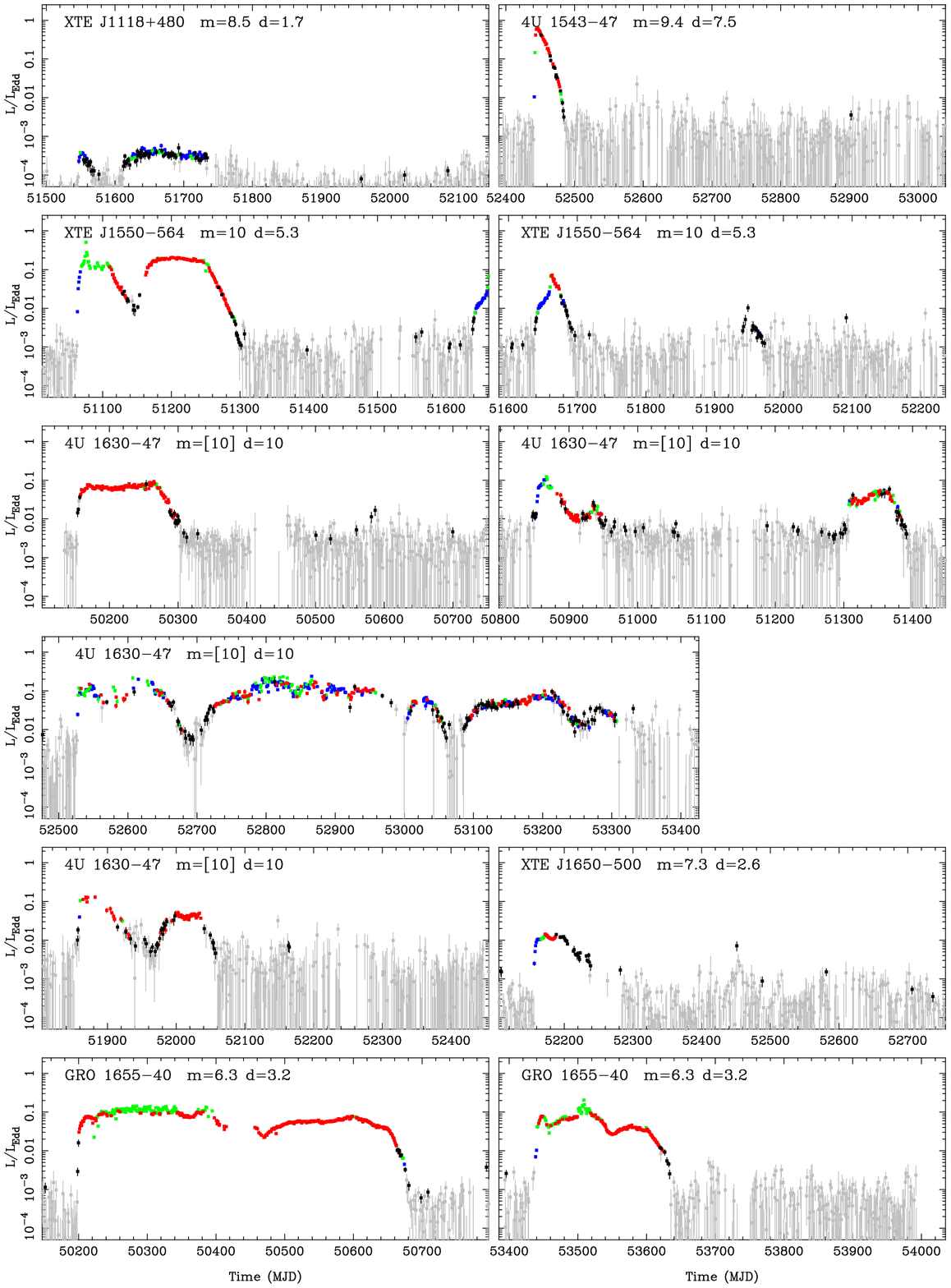}
\end{tabular}
\caption{The long term All Sky Monitor RXTE lightcurves of some
transient BHB (DGK07). In the electronic version these are 
colour coded to indicate their spectral shape (see Fig.~\ref{f:states})
with low/hard in blue, high/soft as red and very
high state in green.
}
\label{f:lc}
\end{figure*}

The mass accretion rate through the entire disc in BHB can vary over
weeks-months-years, triggered by the disk instability (see R. Hynes
in these proceedings or Done, Gierlinski \& Kubota 2007, hereafter 
DGK07). This means that a single object (with constant
distance, inclination and spin) can map out how the spectrum changes
as a function of luminosity as shown in Fig.~\ref{f:lc}.

\begin{figure*}
\begin{tabular}{cc}
\includegraphics[width=0.45\textwidth]{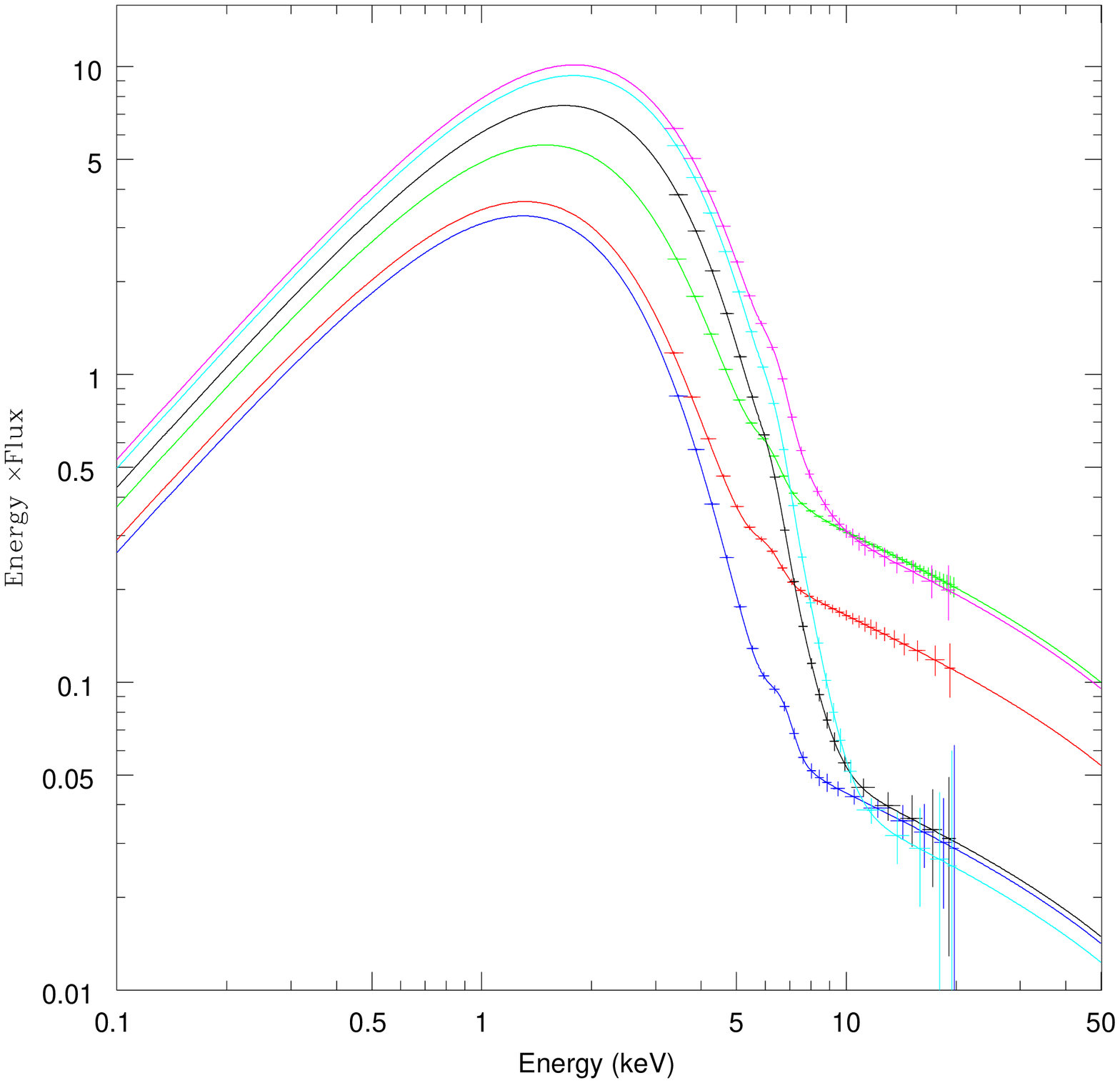}
&
\includegraphics[width=0.45\textwidth]{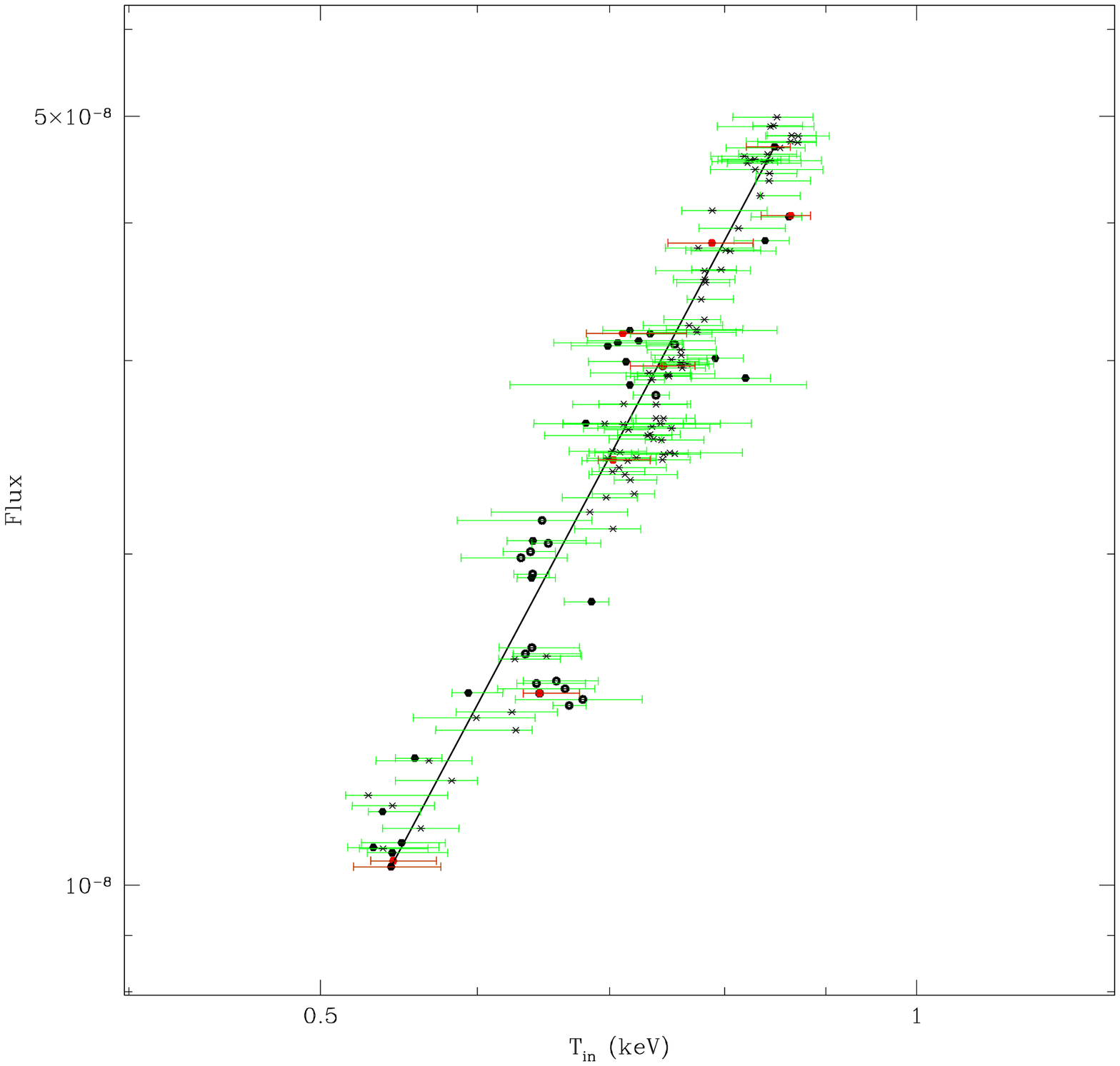}
\end{tabular}
\caption{a) A selection of disc dominated spectra from the transient BHB
GX339-4 b) The disc flux versus its temperature
derived from fits to all the disc dominated data. The line shows the 
$L_{disc}\propto T^4_{max}$ relation expected from a constant size
scale for the inner radius of the disc (Kolehmainen \& Done 2010)}
\label{f:gx339}
\end{figure*}

Fig.~\ref{f:gx339}a shows that these can indeed show spectra which look
very like the simple accretion disc models described above
(section~\ref{s:ss}).  This disc emission also shows very little rapid
variability on timescales of less than a few hundred seconds, as
expected (Churazov et al 2001). Collating disc spectra on longer
timescales at different
$\dot{m}$ gives $L_{disc} \propto T^4_{max}$
(Fig.~\ref{f:gx339}b), clear observational
evidence for a constant size-scale inner radius to the disc despite
the large change in mass accretion rate. This is exactly as predicted
for the behaviour at the last stable orbit, and is a key test of
Einstein's gravity in the strong field limit. Indeed, given how close
the last stable orbit is to the event horizon ($r=6$ compared to the
horizon at $r=2$ for a Schwarzschild black hole), this represents
almost the strongest gravitational field we could ever observe.

Folding in all the required corrections (see above) allows us to
measure this fixed size-scale and translate it into a measure of
spin if there are good system parameter estimates (see R. Hynes 
in these proceedings).
To date, all objects which show convincing $L\propto T^4$ tracks
give moderate spins, as
opposed to extreme Kerr (Davis et al 2006; Shafee et al 2006;
Middleton et al 2006; Gou et al 2010).
This is as expected from current (probably quite
uncertain) supernovae collapse
models (see P. Podsiadlowski, these proceedings, also Gammie, Shapiro
\& McKinney 2004; Kolehmainen \& Done 2010) and BHB in low mass X-ray
binaries should have a spin distribution which accurately reflects
their birth spin. This is because the black hole mass must 
approximately double in order to
significantly change the spin, which is not possible in an LMXB 
even if the black hole (which must be more than $3~M_\odot$)
completely accretes the entire low mass 
($\la 1~M_\odot$) companion star (King \& Kolb 1999).
However, high spins are derived for two objects which do not have
good $L\propto T^4$ tracks, namely GRS1915+105 (McClintock et al 2006, but see 
Middleton et al 2006) and LMC X-1 (Gou et al 2009)

\subsubsection{The high energy tail}

\begin{figure*}[!h]
\begin{tabular}{c}
\includegraphics[width=0.9\textwidth, bb=35 29 370 180,clip=true]{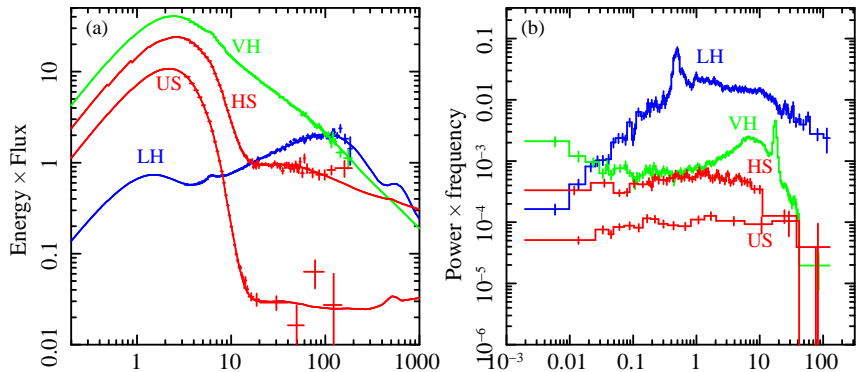}
\end{tabular}
\caption{a) A selection of spectra from the transient BHB GRO
J1655-40. In the electronic version, these different spectral states
are colour coded in the same way as the points in the light-curves
shown in Fig.~\ref{f:lc}. Soft (US and HS: red) spectra are typically
seen at high fractions of Eddington, while hard (LH: blue) spectra are
typically seen at low Eddington fractions. Spectra with both strong
disc and a strong, soft tail (VH: green) are seen at intermediate
luminosities and at the very highest luminosities (DGK07). b) shows
the corresponding variability power spectra from 2-60~keV. The
variability power on these short timescales drops dramatically when
the disc contributes substantially to the spectrum as the disc is very
stable on these timescales, so it dilutes the variability of the
tail.}
\label{f:states}
\end{figure*}

However, even the most disc dominated (also termed high/soft state)
spectra also have a tail extending out to higher energies with $\Gamma
\sim 2$ (see Fig.~\ref{f:gx339}a). This tail carries only a very small
fraction of the power in the data discussed above, but it can be much
stronger and even dominate the energetics.  Where the tail coexists
with a strong disc component (very high/intermediate or steep power
law state) it has $\Gamma> 2$, so the spectra are soft. But where the
disc is weak the tail can dominate the total energy, with $\Gamma< 2$
so the spectra are hard (low/hard state), peaking above 100~keV. All
these very different spectra are shown in Fig.~\ref{f:states}a, colour
coded in the same way as the long term light-curve
(Fig.~\ref{f:lc}). The combination of these two plots shows that 
typically, hard spectra are only seen at low fractions of Eddington,
while disc or disc-plus-soft-tail spectra are seen only at high
fractions of Eddington. This is actually very surprising as
the classic Shakura-Sunyaev disc is unstable to a radiation pressure
instability above $\sim 0.05 L_{Edd}$. Thus we might expect that the
disc is disrupted into some other type of flow at high fractions of
Eddington, and that we see clean disc spectra only at low fractions of
Eddington. This is entirely the reverse of what is seen (as first
recognized by Nowak 1995, see also Gierlinski \& Done 2004a; DGK07).

The tail has more rapid variability than the disc component (Churazov
et al 2001; Gierlinski \& Zdziarski 2005; though the disc 
variability is enhanced on timescales longer than 1~sec 
in the low/hard state: Wilkinson \& Uttley 2009),
typically with large fluctuations on timescales from 100-0.05
seconds. The dynamical (Keplarian orbit) timescale for a $10~M_\odot$
black hole at $6R_g$ is 0.005s, so these timescales are at least $10
\times$ longer than Keplarian.  If the variability is driven by
changes in mass accretion rate then the expected timescale is a factor
$\alpha (H/R)^2$ longer, so this requires $H/R \sim 1$ i.e. a
geometrically thick flow.

\subsection{Theory of geometrically thick flows: ADAFs}

Pressure forces must be important in a geometrically thick flow, by
contrast to a Shakura-Sunyaev disc. If these are from gas pressure
then the flow must be hot, with protons close to the virial
temperature of $10^{12}$~K.  The data require that the electrons are
hot in the low/hard state in order to produce the high energy tail via
Compton scattering (see section \ref{s:comp}), but generally only out
to 100~keV i.e.  $10^9$~K. These two very different requirements on
the temperature can both be satisfied in a two temperature plasma,
where the ion temperature is much hotter than the electron
temperature. This happens in plasmas which are not very dense, as the
electrons can radiate much more efficiently than the protons, so the
electrons lose energy much more rapidly than the protons. Even if the
electrons and protons are heated at the same rate then the proton
temperature will be hotter than the electrons if the protons and
electrons do not interact enough to equilibriate their
temperatures. This gives a further requirement that the optical depth
of the flow needs to be low.

This leads to the idea of a hot, geometrically thick, optically thin flow 
replacing the cool, geometrically thin, optically thick Shakura-Sunyaev disc. 
The exact structure of this flow is not well known at present - Advection 
Dominated Accretion Flows (ADAFs) are the most well known, but there can also be 
additional effects from convection, winds and the jet  (DGK07).  Ultimately, 
magneto-hydrodynamic simulations 
in full general relativity including radiative cooling are 
probably needed to fully explore the complex properties of these flows  (J. 
Hawley, this volume).

When ADAFs (Narayan \& Yi 1995) were first proposed, a key issue was
how to produce such flow from an originally geometrically thin cool
disc (to the extent that one theoretician said 'turbulence generated
by theorists waving their hands').  However, there is now a mechanism
to do this via an evaporation instability. If the cool disc is in
thermal contact with the hot flow then there is heat conduction
between the two, which can lead to either evaporation of the disc into
the hot flow, or condensation of the hot flow onto the disc. At low
mass accretion rates, evaporation predominates in the inner disc,
giving rise to a radially truncated disc/hot inner flow
geometry (Liu et al 1999;
Rozanska \& Czerny 2000; Mayer \& Pringle 2007). This is
exactly the geometry 
required in the phenomenological truncated disc/hot inner flow 
models described in the 
next section

The hot flow can only exist if the electrons and protons do not
interact often enough to thermalise their energy. This depends on
optical depth, and the flow collapses when $\tau \ga 2-3$ which occurs
at $\dot{m}= 1.3\alpha^2\sim 0.01$ for $\alpha=0.1$ (Esin et al 1997).
This is very close to the luminosity of the transition from soft to hard
spectra seen on the outburst decline (Maccarone 2003), though more complicated
behaviour is seen on the hard to soft transition on the rise
(hysteresis: Miyamoto et al 1995; Yu \& Yan 2009), plausibly due to
the rapid accretion rate changes pushing the system into
non-equilibriuim states (Gladstone, Done \& Gierlinski 2007).

But there are issues. I stress again
that the flow should be more complex than an ADAF as
these do not include other pieces of physics which are known to be
present (convection, winds, the jet, changing advected fraction with
radius: see e.g. DGK07). Indeed, standard ADAF solutions are somewhat too optically
thin and hot to match the observed Compton spectra (Malzac \& Belmont
2009), and this is especially an issue at the lowest $L/L_{Edd}$ where
the observed X-ray spectra are far too smooth to be produced by the
predicted very optically thin flow (Pszota et al 2008). 

\subsection{Truncated disc/hot inner flow models}

These two very different types of solution for the accretion flow can
be put together into the truncated disc/hot inner flow model. At high
fractions of Eddington we typically see strong evidence for a disc
down to the last stable orbit (see Fig.~\ref{f:gx339}).  At low
fractions of Eddington, we can have one of these hot, optically thin
geometrically thick flows. The only way to go from one to the other is
for the disc to move inwards. As it penetrates further into the flow
then more seed photons from the disc are intercepted by the flow so
the Compton spectrum softens (see Section \ref{s:comp}).

\begin{figure*}
\begin{tabular}{c}
\hspace{-24pt}
\includegraphics[width=0.8\textwidth,angle=-90]{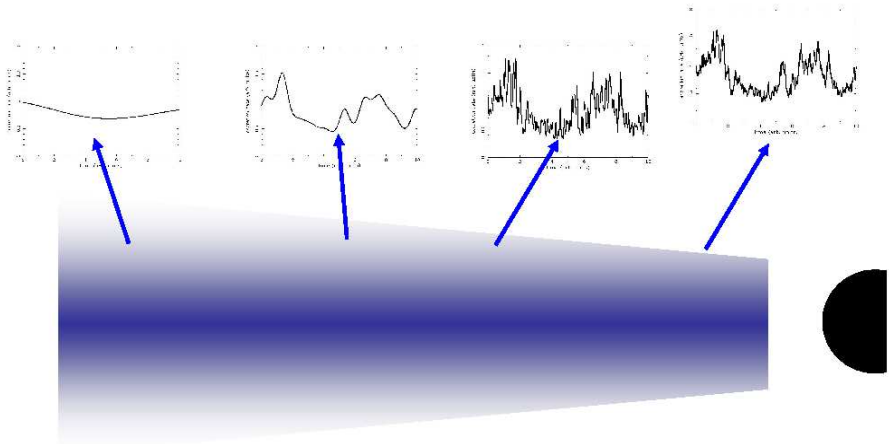}
\end{tabular}
\caption{A propagating fluctuation model. Large
radii produce long timescale fluctuations in mass accretion rate,
which modulate faster variations in mass accretion rate produced by
fluctuations at smaller radii. Figure courtesy of P. Uttley.}
\label{f:fluc}
\end{figure*}

The MRI turbulence in the hot inner flow generates the rapid
variability at each radius, modulating the mass accretion rate to the
next radius. Thus the total variability is the product (not the sum!)
of variability from all radii within the hot flow (Lyubarskii 1997;
Kotov, Churazov \& Gilfanov 2001; Aravelo \& Uttley 2006). This rather
natually gives rise to a key observational requirement that the r.m.s
variability $\sigma$ (see section \ref{s:pow})
in the lightcurve, as measured over a fixed set of frequencies
(duration $T$ and sampling $\Delta t$), is proportional to the mean
intensity $I_0$. This is the r.m.s.-flux relation and cannot be
produced by a superposition (addition) of uncorrelated events such as
the phenomenological 'shot noise' models. Instead this observation
{\em requires} that the fluctuations are multiplicative (Uttley \&
McHardy 2001; Uttley, McHardy \& Vaughan 2005), as sketched in
Fig.~\ref{f:fluc}.

\begin{figure*}
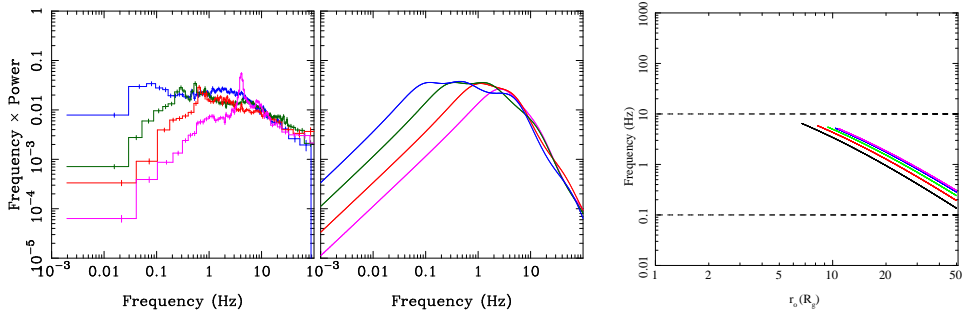

\begin{tabular}{cc}
\includegraphics[width=0.6\textwidth]{pds_trans.ps}
&
\vspace{40pt}
\includegraphics[width=0.36\textwidth]{nuprecro.ps}
\end{tabular}
\caption{a) shows how the power spectra 
evolve as the source makes a
transition from the low/hard to very high state. Low frequency power
is progressively lost as the spectrum softens, while the high frequency
power remains the same. b) shows how this can be modeled in a truncated
disc/hot inner flow geometry, where the tranistion radius decreases,
progressively losing the large radii (low frequency variability) parts of the
hot flow (DGK07). The strong low frequency QPO also moves to higher frequencies,
consistent with Lense-Thirring precession of the hot flow  (Ingram et al 2009).
}
\label{f:pdf_trans}
\end{figure*}

As the disc extends progressively inwards for softer spectra, the flow
at larger radii cannot fluctuate on such large amplitudes as the disc
is underneath it.  The large amplitude fluctuations can only be
produced from radii inwards of the truncated disc. This gets
progressively smaller as the disc comes in, so the longer timescale
(lower frequency) fluctuations are progressively lost, so the power
spectrum narrows, with $\nu_b$ increasing while the amount of high
frequency power stays
approximately the same, as seen in the data (see Fig. ~\ref{f:pdf_trans}a)
These models can be made quantatative, and 
can match the major features of the correlated changes in both
the energy spectra and the power spectra as the source makes a
transition from the low/hard to high/soft states 
(Fig. ~\ref{f:pdf_trans}b, Ingram \& Done 2010). 

However, the power
spectrum also contains a characteristic low frequency QPO which also
moves along with $\nu_b$ (Fig. ~\ref{f:pdf_trans}a).
This can be very successfully modeled in both frequency and
spectrum as Lense-Thirring (vertical) precession of the entire hot
inner flow (Ingram, Done \& Fragile
2009), using the same transition radius as required for the
low frequency break in the broad band power spectrum (Ingram \& Done 2010,
see Fig.~\ref{f:pdf_trans}c). 
Since this is a model involving vertical precession, the QPO
should be strongest for more highly inclined sources, as observed
(Schnittman, Homan \& Miller 2006).

The hot flow is also a good candidate for the base of the jet, in
which case the collapse of the hot flow seen in the high/soft state
spectra also triggers the collapse of the jet, as observed (see
R. Fender, this volume).

\subsection{Scale up to AGN}
\label{s:agn}

These models can be easily scaled up to AGN, keeping the same geometry
as a function of $L/L_{Edd}$ but changing the disc temperature as
expected for a super-massive BH. The luminosity scales as M for mass
accretion at the same fraction of Eddington, and the inner radius
scales as M. Thus the emitting area scales as $M^2$ so $T^4 \propto
L/A \propto 1/M$ so disc temperature {\em decreases}, peaking in the UV
rather than soft X-rays. Interstellar absorption in the host galaxy
and in our galaxy effectively screens this emission (see section 
\ref{s:nabs}), so the disc peak
cannot be directly observed in the same way as in BHB. 

The strong UV flux from the disc also excites 
multiple UV line transitions (see section \ref{s:iabs}) 
from any material around the nucleus. This environment is much
less clean than in LMXRB as there 
are many more sources of gas to be illuminated in the
rich environment of a galaxy centre (molecular clouds,
the obscuring torus...), giving strong line emission
from the broad line region (BLR) and narrow line region (NLR). 

Apart from these differences, we should otherwise see the same behaviour in 
terms of the spectral and variability changes as the mass accretion rate changes, 
and in the jet power.  Thus these models predict intrinsic changes in the 
ionizing nuclear spectrum as a function of mass accretion rate as well as 
changes in the observed spectrum due to obscuration. This is in contrast to
the original 'unification models' of AGN in which Seyfert 1 and 2 nuclei
were intrinsically the same, but viewed at different orientations to 
a molecular torus.

There is growing evidence for intrinsic differences in nuclear spectra.  The 
optical emission line ratios can be quite different in {\em unobscured} AGN of 
similar mass e.g. LINERS show different line ratios to Seyfert 1s which are 
different to Narrow Line Seyfert 1s. This is a clear indication that the 
ionizing spectrum is intrinsically different, as expected from their very 
different $L_{bol}/L_{Edd}$ This can be seen directly from compilations of the 
spectral energy distributions (SED) of these different types of AGN.  The 
fraction of power carried by the X-rays drops for increasing $L_{bol}/L_{Edd}$ 
in much the same way as for BHB. The soft tail at high $L/L_{Edd}$ carries a 
smaller fraction of bolometric luminosity so $L_x$ has to be multiplied by a 
larger factor to get $L_{bol}$ (Vasudevan \& Fabian 2007).

The jet should also change with state as in BHB. This gives a clear potential 
explanation for the origin of radio loud/radio quiet dichotomy. This matches 
quite well to the observed radio populations (Koerding, Jester \& Fender 2006), 
but there is a persistent suggestion that this is not all that is required, with 
the most powerful radio jets being found in the most massive AGN (e.g. Dunlop et 
al 2003). Incorporating super-massive black hole growth and its feedback onto 
galaxy formation into the semi-analytic codes to model the growth of structures 
in the Universe may give the answer to this. These show a correlation between 
super-massive black hole mass and mass accretion rate such that largest black 
holes in massive ellipticals are now all accreting in gas poor environments so 
accrete via a hot flow with correspondingly strong radio jet (Fanidakis et al 
2010).

\section{Compton scattering to make the high energy tail}
\label{s:comp}

Compton scattering is just an energy exchange process between the
photon and electron and the energy exchange is completely
analytic. The output photon energy, $\epsilon_{out}$, 
is given by 
$$\epsilon_{out} = {\epsilon_{in} (1- \beta \cos \theta_{ei}) \over 1-\beta
\cos \theta_{eo} +(\epsilon_{in}/ \gamma)(1-\cos \theta_{io}) }$$ where
$\theta_{ei}$, $\theta_{eo}$ and $\theta_{io}$
are the angles between the electron and input photon, electron and
output photon,  and input and output photon, respectively, 
$\gamma=(1-\beta^2)^{-1/2}$ is the electron Lorentz factor so that its
kinetic energy is $E=(\gamma^2-1)^{1/2} m_e c^2$ and $\epsilon=h \nu
/m_ec^2$ is the photon energy relative to the rest mass energy of the
electron. In general, this simply says whichever of the input 
electron and photon has the most
energy shares some of this with the other. 

An electron at rest has $E=0 < \epsilon$. The photon hits the electron
and momentum conservation means that the electron recoils from the
collision, so the photon loses energy in Compton downscattering. If
the photons and electrons are isotropic and $\epsilon_{in}<<1$
then the angle averaged energy
loss is $\epsilon_{out}=\epsilon_{in}/(1+\epsilon_{in}) \approx
\epsilon_{in}(1-\epsilon_{in})$. Thus the
change in energy $\epsilon_{out}-\epsilon_{in}=\Delta \epsilon =
-\epsilon_{in}^2$. Alternatively, for $\epsilon_{in}>>1$ then the photon
loses almost all its energy in the collision.

\subsection{Thermal Compton Upscattering: Theory}
\label{s:thermal}

\begin{figure*}
\vskip -48pt
\begin{tabular}{c}
\includegraphics[width=0.65\textwidth,angle=-90]{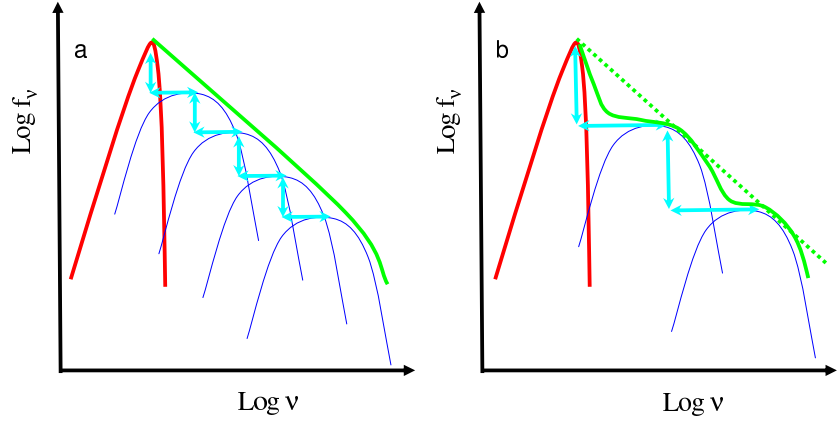}
\end{tabular}
\caption{a) shows how the spectrum built up from
repeated thermal Compton up scattering events for optically thin
($\tau \la 1$) material. A fraction $\tau$ of the seed photons (red)
are boosted in energy by $1+4\Theta$ and then these form the seed
photons for the next scattering, so each scattering order (thin lines:
blue in electronic version) is
shifted down and to the right by the same factor, as indicated by the
arrows (cyan), giving a power law (green solid line). b) shows
that the same spectral index can be obtained by higher $\Theta$ and
lower $\tau$ but the wider separation of the individual scattering
orders result in a bumpy spectrum (green solid line) than a smooth
power law (green dotted line). }
\label{f:thermal}
\end{figure*}

Since Comptonization conserves photon number, it is easiest to draw in
$\log F(E)$ versus $\log E$ as $F(E) d\log E =$~photon number per bin
on a logarithmic energy scale (see section \ref{s:plot}). See also 
the review by Gilfanov (2010). 

In a thermal distribution of electrons, the typical random velocity is
set by the electron temperature $\Theta=kT_e/m_e c^2$ as $v^2 \sim 3
kT_e/m_e$ so $\beta^2=3 \Theta$. Again, for isotropic electon and
photon distributions this can be averaged over angle to give
$\epsilon_{out}=(1+4\Theta+16\Theta^2+...)\epsilon_{in} \approx
(1+4\Theta)\epsilon_{in}$ for $\Theta <<1$. So, in scattering we change
energy by $\epsilon_{out}-\epsilon_{in} = \Delta \epsilon = 4 \Theta
\epsilon_{in}$ and photons are Compton upscattered.  Obviously there
is a limit to this, since the photon cannot gain more energy than the
electron started out with, so this approximation only holds for
$\epsilon_{out} \la 3\Theta$.

Photons can only interact with electrons if they collide. The
probability a photon will meet an electron can be calculated from the
optical depth.  An electron has a (Thomson) cross-section $\sigma_T$
for interaction with a photon.  Thus it is probable that the photon
will interact if there is one electron within the volume swept out by
the photon where the length of the volume is simply the path length,
$R$, and the cross-sectional area is the cross-section for interaction
of the photon with an electron, $\sigma_T$.  Optical depth, $\tau$, is
defined as the number of electrons within this volume, so $\tau=n R
\sigma_T$ where $n$ is the electron number density.  The scattering
probability is $e^{-\tau} \approx 1-\tau$ for $\tau <<1$.

The seed photons are initially at some
energy, $\epsilon_{in}$, so only a fraction, $\tau$, of these are
scattered in optically thin material to energy $\epsilon_{out,1}=
(1+4\Theta)\epsilon_{in}$.  But these scattered photons themselves
also can be scattered to
$\epsilon_{out,2}=(1+4\Theta)\epsilon_{out,1}=(1+4\Theta)^2
\epsilon_{in}$. These photons can be scattered again to
$\epsilon_{out,3}$ etc until they reach the limit of the electron
energy after N scatterings where $\epsilon_{out,N}=(1+4\Theta)^N
\epsilon_{in} \sim 3 \Theta$. Since the energy boost and fraction of
photons scattered is constant then this gives a power law of slope
$\log f(\epsilon) \propto \ln (1/\tau)/\ln (1+4\Theta)$
i.e. $f(\epsilon ) \propto \epsilon^{-\alpha}$ with $\alpha=\ln
\tau/\ln (1+4 \Theta)$. This is a power law from the seed photon
energy at $\epsilon_{in}$ up to $3\Theta$. Thus the power law index is
determined by {\em both} the temperature and optical depth of the
electrons. These cannot be determined independently without
observations at high energy to constrain $\Theta$ as the same spectral
index could be produced by making $\tau$ smaller while increasing
$\Theta$ (see Fig.~\ref{f:thermal}a and
b). However, there are some constraints as the spectrum is only
a smooth power law in the limit where the orders overlap i.e. $\tau$
not too small and $\Theta$ not too big and the energy bandpass
is not close to either the electron temperature or seed photon 
energy (see Fig.~\ref{f:thermal}a and b). 

This list of caveats means that often a power law is {\em not} a good
approximation for a Comptonized spectrum. If the temperatures are non-
relativistic and the optical depth not too small then {\sc comptt}
(Titarchuk 1994) or {\sc nthcomp} (Zdziarski Johnson \& Magdziarz
1996, where the 'n' in front of thcomp denotes that it does not have
the reflected component of Zycki, Done \& Smith 1999) can be used as
these both include the downturn in the Compton emission close to the
seed photon energy which affects the derived disc properties
(Kubota \& Done 2004), as well as the rollover at the electron
temperature. 

\begin{figure*}
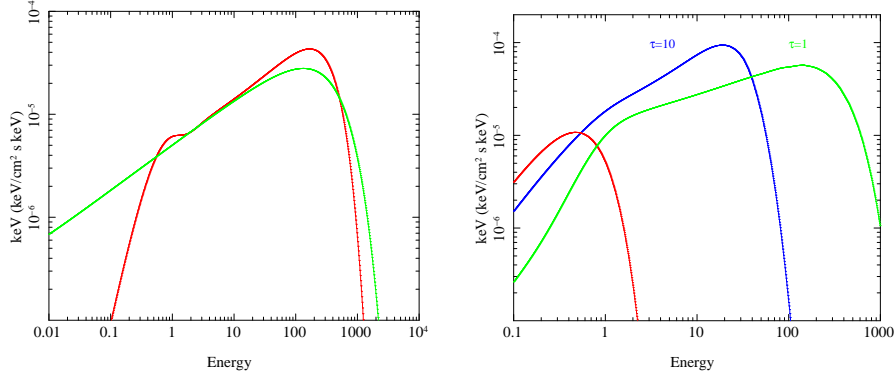

\begin{tabular}{cc}
\includegraphics[width=0.45\textwidth]{exp_comp.ps}
&
\includegraphics[width=0.45\textwidth]{eqpair.ps}
\end{tabular}
\caption{a) The dark grey (red) line shows the themal Comptonization
spectrum computed by {\sc compps} for $\tau=2$ and $kT_e=100$~keV with
seed photons at 0.2~keV (using geom=0 and covering fraction=1).  The
slight bump at the seed photon energy is from the fraction
$exp(-\tau)$ of the seed photons which escape without scattering.  The
Compton upscattered spectrum does not extend down below the seed
photon energy, so there is an abrupt downturn below
$3kT_{seed}=0.6$~keV.  By contrast, an exponential cutoff power law
spectrum (light grey/green) has a much more gradual rollover at the
electron temperature, and extends below the seed photons. These two
differences can mean that an exponentially cutoff power law is a poor
approximation to a Comptonised spectrum.  b) shows an energetic
approach to Comptonisation for $\ell_h/\ell_s=10$ with $\tau=10$ (dark
grey/blue) and $1$ (light grey/green). Each of these two Comptonised
spectra contains $10\times$ more energy than the seed photons (thermal
spectrum at low energies: red), but have different spectral index and
electron temperature.}
\label{f:xspec_comp}
\end{figure*}

However, for temperatures much above $\sim 100$~keV with
good high energy data then relativistic corrections become important
and {\sc compps} (Poutanen \& Svensson 1996) or {\sc eqpair} (Coppi
1999) should be used.  The Compton rollover at the electron
temperature is rather sharper than an exponential, so using an
exponentially cutoff power law is not a good approxmation (see
Fig.~\ref{f:xspec_comp}a), and will distort the 
derived reflected fraction (see Section~\ref{s:refl}).

\begin{figure*}
\vskip -48pt
\begin{tabular}{c}
\includegraphics[width=0.65\textwidth,angle=-90]{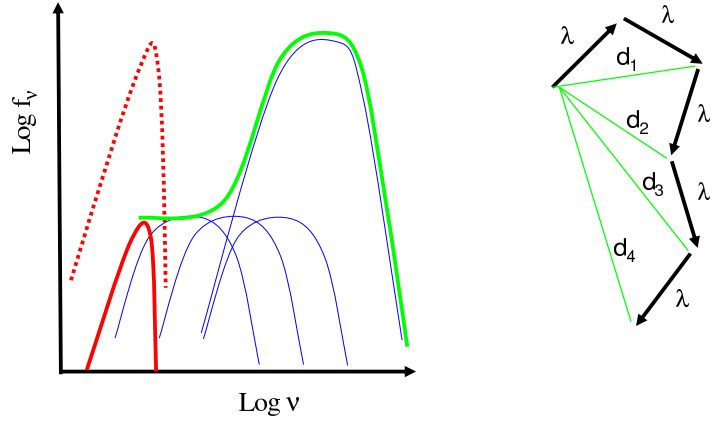}
\end{tabular}
\caption{The left hand panel shows how the spectrum built up from
repeated thermal Compton up scattering events for optically thick
($\tau \ga 1$) material.  Almost all ($\exp^{-\tau}$) the original
seed photons (red dotted line)
are scattered multiple times until they all pile up at the mean
electron energy of $\sim 3\Theta$. The right hand sketch shows the
electron path, randomising direction by scattering after an average
distance $\lambda=1/n\sigma_T$ (i.e. travelling $\tau=1$).  }
\label{f:optthick}
\end{figure*}

In optically thick Comptonization, with $\tau\gg 1$, almost all the
photons are scattered each time so almost all of them end up at the
electron temperature of $3\Theta$ forming a Wien peak
(Fig.~\ref{f:optthick}a).  The average distance a photon travels before
scattering is $\tau=1$ i.e. a mean free path of $\lambda=1/(n
\sigma_T)$. Thus after 1 scattering, the distance travelled is
$d_1^2=\lambda^2 +\lambda^2 - 2 \lambda^2 \cos \theta_{12}$ while
after 2 scatterings this is $d_2^2=d_1^2 + \lambda^2 - 2 \lambda d_1
\cos \theta_{2,3}$ and after N scatterings $d_N^2=d_{N-1}^2 +
\lambda^2-2 \lambda d_{N-1} \cos \theta_{N,N+1}$.  Since the
scattering randomizes the direction then the angles average out,
leaving $d_N^2=N \lambda^2= N /(n \sigma_T)^2$. The photon can escape
when $d_N=R$, so the average number of scatterings before escape is
$N\sim \tau^2$ (see Fig.~\ref{f:optthick}b).

The amount of energy exchange from the electrons to photons in
Comptonization can be roughly characterized by the Compton $y$
parameter. The fractional change in energy of the photon distribution,
$y$, is the average number of scatterings times average fractional
energy boost per scattering such that $y \approx (4\Theta + 16
\Theta^2)(\tau+\tau^2) \approx 4\Theta \tau^2$ in the optically thick,
low temperature limit. If $y\ll 1$ then the electrons make very little
difference to the spectrum, while for $y \ga 1$, Comptonization is
very important in determining the emergent spectrum.

\subsection{Comptonization via energetics}

Describing the spectrum in terms of $\tau$ and $\Theta$ is the
'classical' way to talk about Compton scattering. But the physical
situation is better described by $\tau$ and energetics. There is some
electron region with optical depth $\tau$, heated by a power input
$\ell_h$, making a Comptonized spectrum from some seed photon
luminosity $\ell_s$. Now we are doing this by luminosity instead of
photon number we should
plot $\nu F(\nu)$ rather than $F(\nu )$. 
The seed photons peak at $\sim 3kT_{seed}$, with $\nu F(\nu)=\ell_s$
The 'power law' 
Compton spectrum always points back to this point, forming a power law
of energy
index $\alpha\sim \log\tau/\log(1+4\Theta)$ which extends from here to 
$\sim 3kT_e$, and has total power $\ell_h$.
The resulting equation can then be solved for
$\Theta$ (see Haardt \& Maraschi 1993). This is shown in 
Fig.~\ref{f:xspec_comp}b, using the {\sc eqpair}
model. The seed photons (red) are Comptonised
by hot electrons which have $10\times$ as much power as in the seed photons.
For a large optical depth (blue: $\tau=10$) this energy is shared between many 
particles, so the electron temperature is lower than for 
a smaller optical depth (green: $\tau=1$). The spectrum cannot extend
out to high energies, so has to be harder in order to contain the requisite
amount of power. 

This energetic approach gives
more physical insight when we come to consider seed photons produced
by reprocessing of the hard X-ray photons illuminating the disc (see
section ~\ref{s:rep}).

\subsection{Thermal Compton Scattering: Observations of Low/Hard
 State}
\label{s:1118}

\begin{figure*}
\begin{tabular}{cc}
\includegraphics[width=0.45\textwidth]{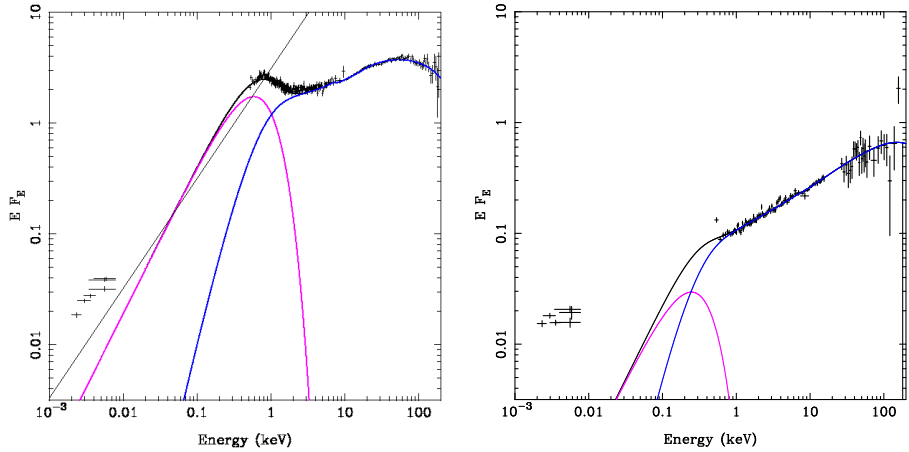}
&
\includegraphics[width=0.45\textwidth]{1753dim.ps}
\end{tabular}
\caption{ The unabsorbed data from the BHB transient 1753.5-0127
modelled with a disk (thermal compenent peaking at 0.6~keV: magenta)
and thermal Compton scattering of seed photons from the disc (peaking
at 80~keV: blue). The left hand panel shows a bright low/hard state,
where the optical/UV data clearly lie far below the extrapolated hard
X-ray emission, indicating that the X-ray spectrum must break
(i.e. have its seed photons) somewhere between the UV and soft X-ray
bandpass. The disc is then the obvious source for these seed
photons. The optical/UV points are some way below the extrapolated
disc emission, but this leaves room for hard X-ray reprocessing, and
possibly some component from the radio jet spectrum extrapolated up to
this band (thin black line).  However, the right panel shows the very
different dim low/hard state spectrum from the same source. There is
still a very weak soft X-ray component, but the optical/UV now lie on
the extrapolated hard X-ray spectrum, making it more likely that the
seed photons for the Compton scattering are at energies below the
optical, i.e. are probably cyclo-synchrotron (Chiang et al 2009)}
\label{f:1753}
\end{figure*}

Fig.~\ref{f:1753} shows two examples of low/hard state spectra from a
BHB, one which rolls over at $\sim 90$~keV, and one which extends to
$\sim 200$~keV. Both rollovers look like thermal Comptonization, but
with different temperatures ($\sim 30$~keV, and $\sim 100$~keV,
respectively, i.e. $\Theta \sim 0.06$ and $0.2$).  The optical depth
can then be derived from the spectral index but not quite as easily as
described above as $\tau$ is of order unity rather than $\ll 1$ as
required for the analytic expression. A more proper treatment gives
$\tau \sim 0.6$.  Then the fraction of unscattered seed photons should
be only $1-e^{-\tau}$ but we actually see more than this in
Fig.~\ref{f:1753}a. Thus we require that not all the seed photons go
through the hot electron region i.e. the geometry is either a
truncated disc or the electron regions are small compared to the disc
- either localised magnetic reconnection regions above the disc or a
jet. See Section~\ref{s:rep} for how the energetics of Compton
scattering strongly favour the truncated disc.

The spectra shown above are fit assuming that the observed soft X-ray
component provides the seed photons for the Compton upscattering.
This can be seen explicitly where there are multiwavelength
observations, extending the bandpass down to the optical/UV where the
outer parts of the disc can dominate the emission. Fig.~\ref{f:1753}a
shows this for a bright low/hard state in the transient BHB XTE
J1753.5-0127 (Chiang et al 2009). It is clear that extrapolating the hard X-ray power law
down to the optical/UV will produce far more emission than
observed. Thus the hard X-ray power law must break between the UV and
soft X-ray, i.e. the seed photons for the Compton upscattering should
be somewhere in this range. Since there is an obvious soft X-ray
thermal component, this is the obvious seed photon
identification. Fitting the soft X-rays with a disc component slightly
underproduces the optical/UV emission, but this can be enhanced by
reprocessing of hard X-rays illuminating the outer disc (van Paradijs
1996).

However, for dimmer low/hard states, the hard X-rays extrapolate
directly onto the optical/UV emission (Fig.~\ref{f:1753}b: Chiang et
al 2009; Motch et al 1985). This
looks much more like the seed photons are at lower energies than the
optical. In the truncated disc picture, the disc can be so far away
that it subtends a very small solid angle to the hot electron region
which is concentrated at small radii. Thus the amount of seed photons 
from the disc illuminating the hot electrons
can be very small, and can be less important than 
seed photons produced by the hot flow itself.
The same thermal electrons as make the Compton
spectrum can make both bremsstrahlung (from interactions with protons)
and cyclo-synchrotron (from interactions with any magnetic field such
as the tangled field produced by the MRI). The bremsstrahlung spectrum
will peak at $kT_e$, so these seed photons have similar energies to the
electrons so cannot gain much from Compton scattering. However, the
cyclo-synchroton typically peak in the IR/optical region so these can
be the seed photons for a power law which extends from the optical to
the hard X-ray region (Narayan \& Yi 1995; Di Matteo, Celotti \&
Fabian 1997; Wardzinski \& Zdziarski 2000; Malzac \& Belmont 2009).

Evidence for a change in seed photons is also seen in the variability
(see R. Hynes, this volume). In bright states (both bright low/hard
and high/soft/very high states) the optical variability is a lagged
and smoothed version of the X-ray variability, showing that it is from
reprocessed hard X-ray illumination of the outer accretion
disc. However, this changes in the dim low/hard state, with the
optical having more rapid variability than the hard X-rays, and often
{\em leading} the hard X-ray variability (Kanbach et al 2001; Gandhi et al
2008; Durant et al 2008; Hynes et al 2009). This completely rules out
a reprocessing origin, clearly showing the change in the optical
emission mechanism. 

\subsection{Non-Thermal Compton Scattering: High/Soft State}
\label{s:nonthermal}

\begin{figure*}
\begin{tabular}{cc}
\includegraphics[width=0.45\textwidth]{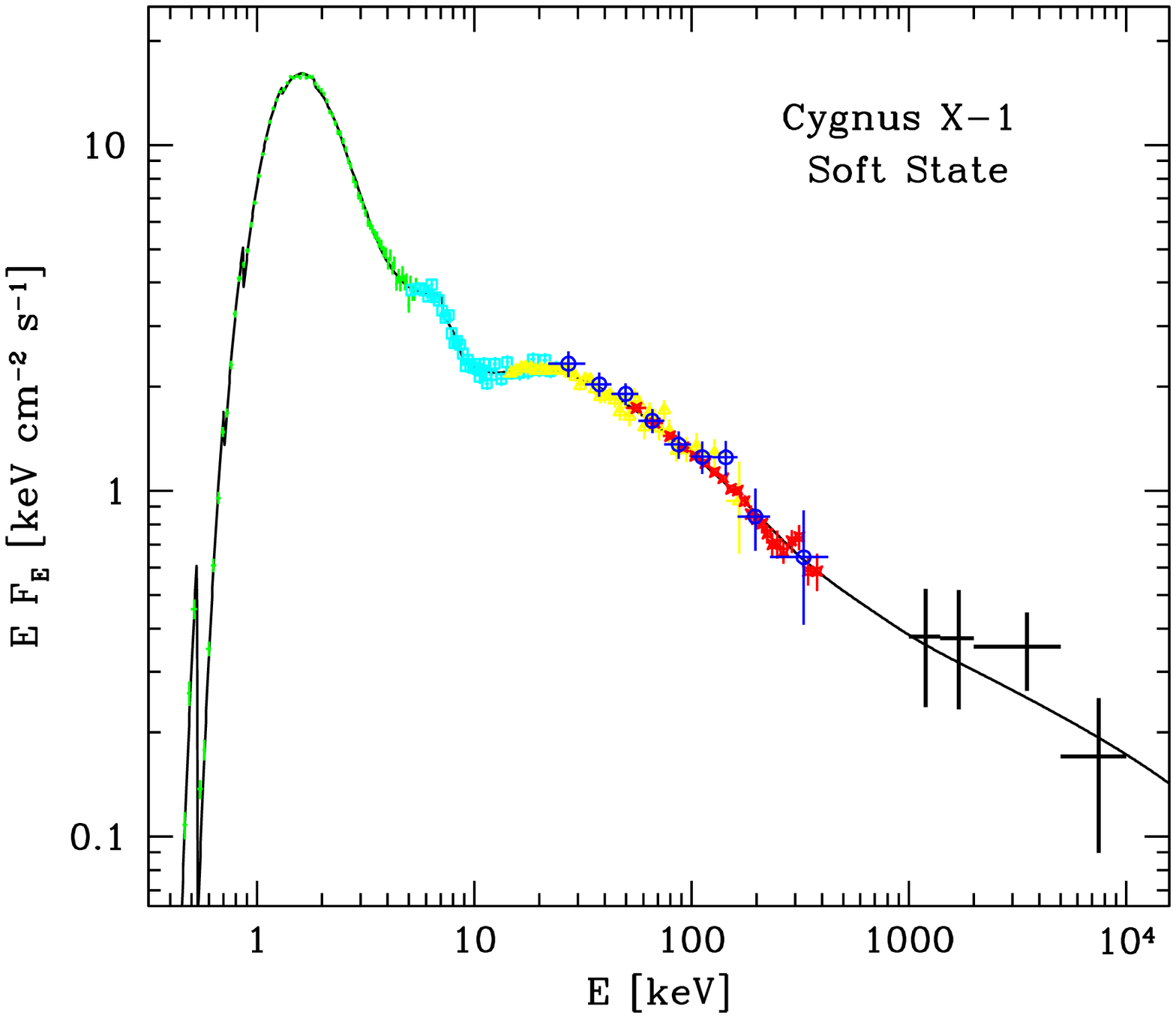}
&
\includegraphics[width=0.45\textwidth,angle=90]{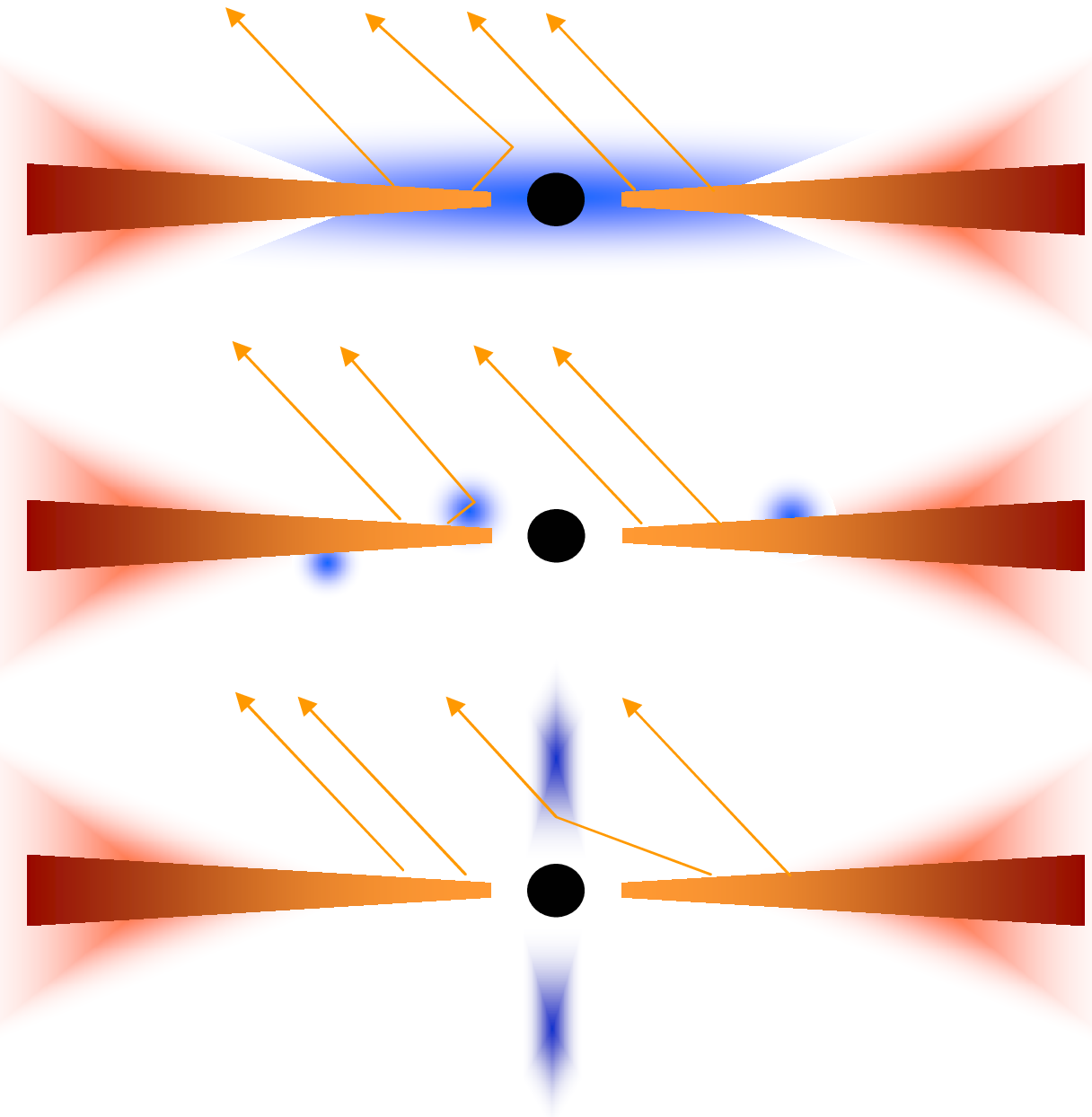}
\end{tabular}
\vskip -36pt
\caption{The left panel shows a composite spectrum of the high/soft
state of Cyg X-1 from multiple instruments. Interstellar absorption is
not removed, causing the drop below 1~keV. Nonetheless, the disc
clearly dominates the low energy data (BeppoSAX LECS: green) while the
soft tail extends out to $\sim 10$~MeV (OSSE COMPTEL: black). The tail
itself is smooth, but there is curvature and spectral features seen
from 5-20~keV (BeppoSAX HPGSPC: cyan) from reflection (see
Section~\ref{s:refl}). Figure from McConnell et al (2002). The right
panel shows the potential source geometries in which the disc photons
can dominate the spectrum at low energies as observed. This requires
that either the electron acceleration region is optically thin, or
that it is localized, perhaps as active regions 
over the disc or in the (base of the)
jet.}
\label{f:cygx1_soft}
\end{figure*}

While the low/hard state can be fairly well described by thermal
Compton scattering, the same is not true for the tail seen in the high
soft state. This has $\Gamma \sim 2$ and clearly extends out past
1~MeV, and probably past 10~MeV for Cyg X-1 (Gierlinski et al 1999;
McConnell et al 2002) , as shown in Fig.~\ref{f:cygx1_soft}a. If
this were thermal Compton scattering then the electron temperature
must be $\Theta \ga 1$, requiring optical depth $\tau <<1$ in order to
produce this photon index. The separate Compton orders are then well
separated and the spectrum should be bumpy (see Fig.~\ref{f:thermal}b)
rather than the smooth power law seen in the data (the bump in the
data is from reflection: see Section~\ref{s:refl}). 
Thus this tail cannot be
produced by thermal Comptonization. 

Instead it can be non-thermal
Compton scattering, where the electron number density has a power law
distribution rather than a Maxwellian i.e. $n(\gamma) \propto
\gamma^{-p}$ from $\gamma=1$ to $\gamma_{max}$.  Going back to the
original equation for the Compton scattered energy boost, for
$\gamma>>1$ the output photon is beamed into a cone of angle
$1/\gamma$ along the input electron direction. This gives an angle
averaged output photon energy of $\epsilon_{out}= (4/3 \gamma^2
-1)\epsilon_{in} \approx \gamma^2 \epsilon_{in}$ for an isotropic
distribution of input photons and electrons. Thus the Compton
scattered spectrum extends from $\epsilon_{in}$ to $\gamma^2_{max}
\epsilon_{in}$, forming a power law from a single scattering order.

The power law index of the resulting photon spectrum can be calculated
from an energetic argument. The rate at which the electrons lose
energy is the rate at which the photons gain energy, giving
$F(\epsilon ) d\epsilon \propto \dot{\gamma} n(\gamma) d\gamma$ where
$\dot{\gamma}\propto \gamma^2$ is the rate at which a single electron
of energy $\gamma$ loses energy and $n(\gamma)$ is the number of
electrons at that energy. Thus $F(\epsilon) \propto \gamma^2
\gamma^{-p} d\gamma/d\epsilon$. Since $\epsilon\sim
\gamma^2\epsilon_i$ then $d\epsilon/d\gamma=2\gamma$ so $F(\epsilon)
\propto \gamma^{-(p-1)} \propto \epsilon^{-(p-1)/2}$ i.e. an energy
spectral index of $\alpha=(p-1)/2$ (G. Ghisellini, private communication).

\begin{figure*}
\vskip -80pt
\begin{tabular}{l}
\includegraphics[width=0.7\textwidth,angle=-90]{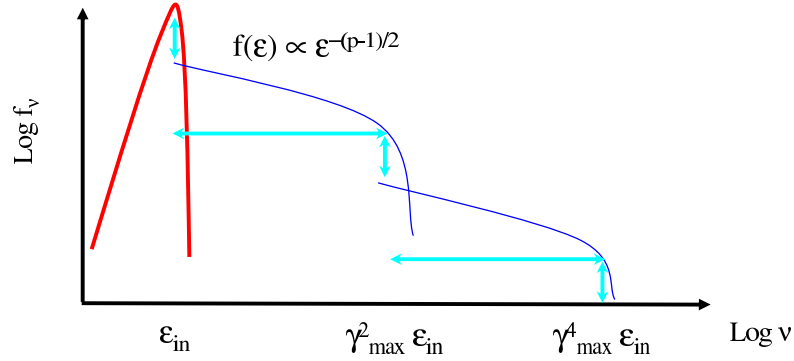}
\end{tabular}
\caption{Schematic of non-thermal Compton scattering, where the seed
  photons (red) form a power law from a single scattering due to the
  power law electron distribution from $\gamma=1-\gamma_{max}$. 
}
\label{f:nonthermal}
\end{figure*}

For an optically thin electron region, the electrons intercept only a
fraction $\tau$ of the seed photons, and scatter them to
$\gamma^2_{max} \epsilon_{in}$ with an energy spectral index of
$(p-1)/2$. These can themselves be scattered into a second order
Compton spectrum to $\gamma^2_{max} (\gamma^2 \epsilon_{in}) = \gamma^4
\epsilon_{in}$ but very soon the large energy boost means that these
hit the limit of the electron energy of
$\epsilon_{out}=\gamma_{max}$. The resulting spectrum is shown
schematically in Fig.~\ref{f:nonthermal}.

Thus the high/soft state of Cyg X-1
requires $\gamma_{max} > 30$ to get seed photons from the disc at
1~keV upscattered to 1~MeV, or  $\gamma_{max} > 100$ to get to 10~MeV,
while the energy spectral index of $1.2 =
(p-1)/2$ implies $p\sim 3$. Such non-thermal Compton spectra can be
modelled using either {\sc compps} or {\sc eqpair}.

Fig.~\ref{f:cygx1_soft}a shows that the seed photons from the disc
are clearly seen as distinct from the tail. This requires that either
the optical depth is very low, or the electron acceleration region does not
intercept many of the seed photons from the disc i.e. localized
acceleration regions as shown schematically in Fig.~\ref{f:cygx1_soft}b.

\subsection{Thermal - Nonthermal (Hybrid) Compton Scattering}
\label{s:hybrid}

\begin{figure*}
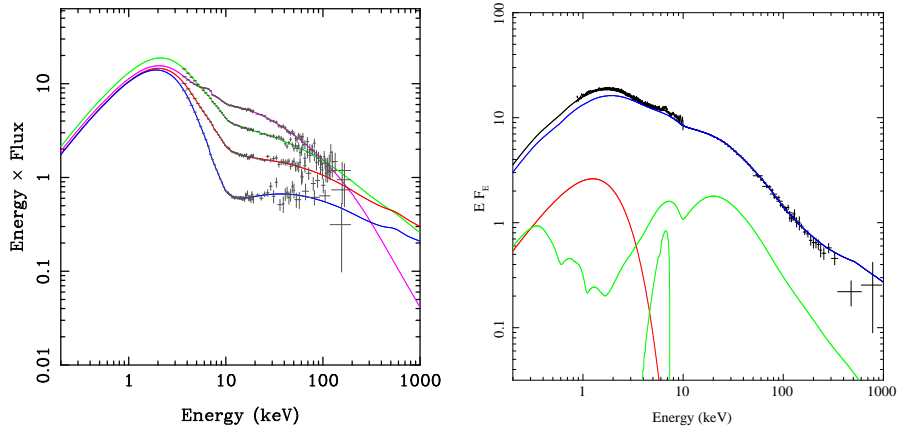
 \begin{tabular}{cc} 
\includegraphics[width=0.45\textwidth]{trans_vth.ps} & 
\includegraphics[width=0.45\textwidth]{1550_2.ps} 
\end{tabular} 
\caption{a) Transition from high/soft to extreme very high state in
the BHB transient XTE~J1550-564. b) Broader band spectrum of the
extreme very high state. The emission is dominated at all energies by
the Comptonized continuum, and there is very little of the disc
emission (thermal component peaking at $\sim 1$~keV: red) that can be
seen directly. This requires that the corona is optically thick and
covers most of the inner disc, in contrast to the possible high/soft
state geometries.  The reflected continuum and its associated iron
emission line (light grey/green) are also shown (see section
\ref{s:refl}).}
\label{f:vhs} 
\end{figure*}

The high/soft states can transition smoothly into the very high or
intermediate state spectra, with the tail becoming softer and
carrying a larger fraction of the total power (Fig.~\ref{f:vhs}a). The
disc then merges smoothly into the tail, showing that the hot electron
region completely covers the inner disc emission and it is optically
thick. The tail still extends up to 1~MeV, so clearly also contains
non-thermal electrons, and is rather soft but has a complex curvature
(see Fig.~\ref{f:vhs}b).

\begin{figure*} 
\vskip -50pt
\begin{tabular}{l} 
\includegraphics[width=0.7\textwidth,angle=-90]{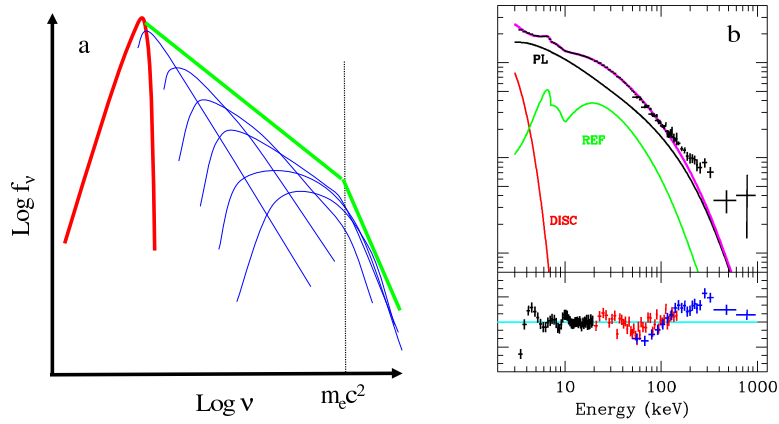} 
\end{tabular} 
\caption{ a) Spectrum resulting from a steep power law electron
distribution which is optically thick ($\tau\ga 1$ rather than
$\tau\gg 1$). The steep power law means that the mean electron energy
is low, so the spectrum is built up from multiple Compton scatterings
in a way similar to thermal Comptonization. Thus $\epsilon\gamma\sim
1$ at $\epsilon\sim 1$ i.e. at 511~keV, so the spectrum breaks due to
the reduction in cross-section. b) shows how this means that such a
steep power law electron distribution cannot fit the high energy tail
seen in the very high state.}
\label{f:steep} 
\end{figure*}

The extent of the tail shows that there must be non-thermal Compton
scattering, as in the high/soft state. However, the tail is
softer so the electron index must be more negative
than in the high/soft state spectrum, so the mean electron
energy is lower.  Yet the lack of direct disc emission requires an 
optical depth of $ \ga 1$. This means that there
are multiple Compton scattering orders forming the spectrum in a
similar way to thermal Compton scattering, but from a non-thermal
distribution. The energetic limit to which photons can be scattered
is $\gamma_{max}$ but because the energy boost on each scattering is
small, the spectrum actually rolls over at $m_ec^2=511$~keV as the
cross-section for scattering drops at this point where
$\gamma\epsilon\sim 1$ (as the cross-section transitions to
Klein-Nishina rather than the constant Thomson cross-section seen at
lower collision energies). Thus optically thick, nonthermal
Comptonized spectra with a steep power law electron distribution 
does not produce a power law spectrum. Instead there is a
break at 511 keV (Ghisellini 1989), as shown schematically in
Fig.~\ref{f:steep}a, which means that this 
cannot fit the observed tail at high energies 
seen in 
the very high state spectrum, as shown in Fig.~\ref{f:steep}b
(Gierlinski \& Done 2003)

Thus neither thermal nor non-thermal Compton scattering can produce
the tail seen in these very high state data. Instead, the spectra
require both thermal and non-thermal electrons to be present. This
could be produced in a single acceleration region, where the initial
acceleration process makes a non-thermal distribution but where the
resulting electrons have a hybrid distribution due to lower energy
electrons predominantly cooling through Coulomb collisions (which
thermalize) while the higher energy electrons maintain a power law
shape by cooling via Compton scattering (Coppi 1999). Such hybrid
thermal/non-thermal spectra can be modeled in {\sc xspec} using
either {\sc compps} or {\sc eqpair}.  Alternatively, this could
indicate that there are two separate acceleration regions, one with
thermal electrons, perhaps the remnant of the hot inner flow, and one
with non-thermal, perhaps magnetic reconnection regions above the disc
or the jet (DGK07). This could be modeled by two separate {\sc compps} or {\sc
eqpair} components, one of which is set to be thermal, and the other
set to be non-thermal.

Whatever the electrons are doing, they are optically thick and cover
the inner disc. Hence it is very difficult to reconstruct the
intrinsic disc spectrum in these states. The derived temperature and
luminosity of the disc depend on how the tail is modeled. A simple
power law model for the tail means that it extends below the putative
seed photons from the disc. Instead, in Compton upscattering, the continuum
rolls over at the seed photon energy, so there are fewer photons
at low energies from the tail so the disc has to be more luminous
and/or hotter in order to match the data. Compton scattering conserves
photon number, so all the photons in the tail were initially part of the 
disc emission, so the intrinsic disc emission is brighter than that observed by
this (geometry dependent) factor. This means that the  
intrinsic disc luminosity and temperature 
cannot be unambiguously recovered from the data 
when $\tau\ga 1$, where the majority of
photons from the disc are scattered into the tail (Kubota et al 2001; 
Kubota \& Done 2004; Done \& Kubota 2006; Steiner et al 2009). 

For more observational details of spectral states see
Remillard \& McClintock (2006) and Belloni (2009).

\section{Atomic Absorption}
\label{s:abs}

The intrinsic continuum is modified by absorption by material along
the line of sight. This can be the interstellar medium in our galaxy
the host galaxy of the X-ray source, or material associated with the
source. For AGN this can be the molecular torus, the NLR or BLR
clouds,  or an accretion disc wind. For BHB this is just an accretion
disc wind, and any wind from the companion star.  For magnetically
truncated accretion discs such as seen in the intermediate polars
(white dwarf) and accretion powered millisecond pulsars (neutron
stars) there is an accretion curtain which can be in the line of sight
(de Martino et al 2004), while for extreme magnetic fields the disc is
completely truncated and there is only an accretion column (polars in
white dwarfs) but this overlays the X-ray hot shock, giving complex
absorption (Done \& Magdziarz 1998).

Again the key concept is optical depth, $\tau = \sigma(E) n R$, only
now the cross-section, $\sigma(E)$, has a complex dependence on energy
(rather than the constant electron scattering cross-section,
$\sigma_T$, for energies below 511~keV).
We can combine $nR =N_H$ as the number of Hydrogen atoms along a line
of sight volume with cross-sectional area of 1~cm$^{-2}$, so the optical depth 
is simply related to column density by 
$\tau(E) = \sigma(E) N_H$. 

\subsection{Neutral Absorption}
\label{s:nabs}

The photo-electric absorption cross-section of neutral hydrogen is
zero below the threshold energy of 13.6~eV, below which the photons do
not have enough energy to eject the electron from the atom. It peaks
at this threshold edge energy at a value of $6\times
10^{-18}$~cm$^{-2}$ and then declines as $ \approx
(E/E_{edge})^{-3}$. Thus the optical depth is unity for a Hydrogen
column of $1.6\times 10^{17}$ cm$^{-2}$ at 13.6~eV while a typical
column through our galaxy is $>10^{20}$~cm$^{-2}$, showing how
effectively the UV emission is attenuated.

\begin{figure*}
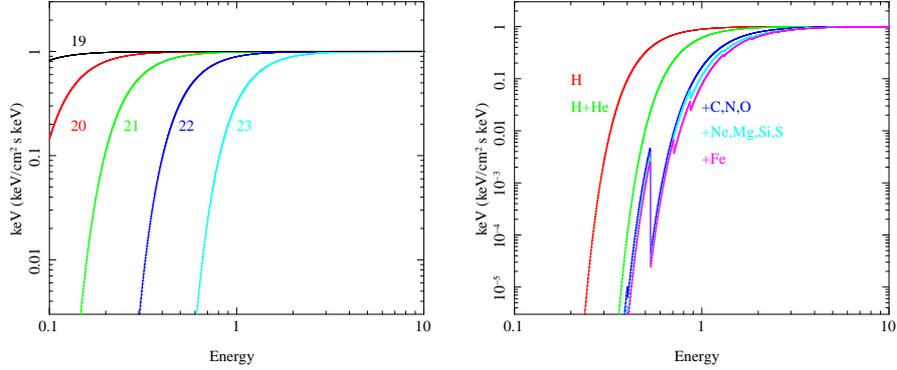
 
\begin{tabular}{cc} 
\includegraphics[width=0.45\textwidth]{nh.ps} 
& 
\includegraphics[width=0.45\textwidth]{element.ps}\\ 
\end{tabular} 
\caption{ a) Photo-electric absorption from Hydrogen alone, in a
column of $\log N_H=19,20,21,22,23$ (from left to right). b) shows how
the absorption for a column of $\log N_H=22$ changes as progressively heavier 
atomic numbers are added.} 
\label{f:col} 
\end{figure*}

The drop in cross-section with energy means that an H column of
$10^{20}$~cm$^{-2}$ has $\tau=1$ at an energy of $\sim 0.2$~keV, allowing the soft X-rays to
be observed. Fig.~\ref{f:col}a shows how a 
hydrogen column of $\log N_H=19,20,21,22,23$ progressively 
absorbs higher energy X-rays.  
However, the column is not made up soley of H. There are
other, heavier, elements as well.  
These have more bound electrons, but the highest edge
energy will be from the inner $n=1$ shell (also termed the K shell)
electrons as these are the closest to the nuclear charge. Since this charge is
higher, then the $n=1$ electrons are more tightly bound than those
of H  e.g. for He it is 0.024~keV, C, N, O is 0.28, 0.40 and
0.53~keV. These elements are less abundant than H so they form small
increases in the total cross-section.  Fe is the last astrophysically
abundant element, and this has a K edge energy of 7.1~keV. Fig.~\ref{f:col}b
shows this for column of $10^{22}$~cm$^{-2}$ for progressively adding higher atomic 
number elements assuming solar abundances. Helium has an impact on the total
cross-section, but additional edges from heavier elements are important
contributions to the total X-ray absorption, especially Oxygen.

In {\sc xspec}, this can be modeled using {\sc tbabs}/{\sc zthabs}
(Wilms, Allen and McCray 2000, where the latter has redshift as a free
parameter) or {\sc phabs}/{\sc zphabs} (Balucinska-Church \& McCammon
1992) if the abundances are assumed to be solar, or {\sc
tbvarabs}/{\sc zvphabs} if the data are good enough for the individual
element abundances to be constrained via their edges such as in
GRS~1915+104 (Lee et al 2002). However, with excellent spectral
resolution data from gratings then the line absorption (especially
from neutral Oxygen) becomes important (see section~\ref{s:lines}),
and {\sc tbnew} (see J. Wilms web page at
http://pulsar.sternwarte.uni-erlangen.de/wilms/research/tbabs/) should
be used (Juett, Schulz \& Chakrabarty 2004).

\subsection{Ionised Absorption}
\label{s:iabs}

Photo-electric absorption leaves an ion i.e. the nuclear charge is not
balanced.  Thus all the remaining electrons are slightly more tightly
bound, so all the energy levels increase. The ion can recombine with
any free electrons, but if the X-ray irradiation is intense then the
ion can meet an X-ray photon before it recombines, so that the
absorption is dominated by photo-ionized ions. For H, this means there
is no photo-electric absorption, since it has no bound electrons after
an ionization event, so some fraction of the total cross-section
disappears. Helium may then have 1 electron left, so its edge moves to
0.052~keV.  At higher ionizations, Helium is completely ionized, so its
contribution to the total cross-section is lost and there are only
edges from (ionized) C, N, O and higher atomic number elements. Thus
the effect of going to higher ionization states is to reduce the
overall cross-section as the numbers of bound electrons are lower.  In
the limit where all the elements are completely ionized, there is no
photo-electric absorption at all.

\begin{figure*}
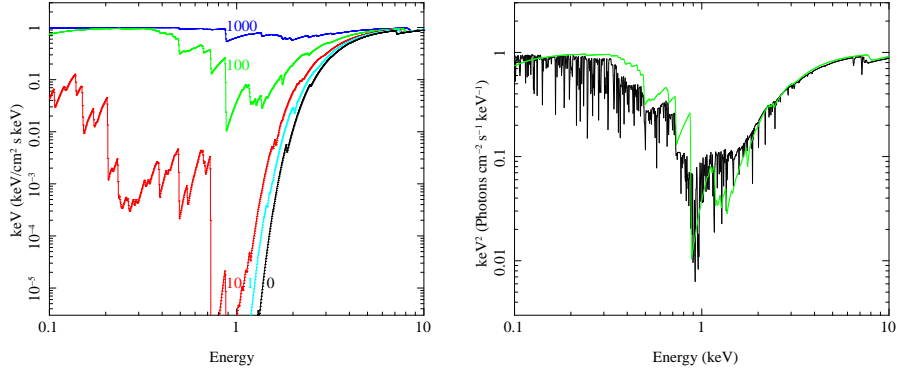
 
\begin{tabular}{cc} 
\includegraphics[width=0.45\textwidth]{absori.ps} 
& 
\includegraphics[width=0.45\textwidth]{xstar.ps}\\ 
\end{tabular} 
\caption{ a) The photo-electric (bound-free) absorption of a 
column of $N_H=10^{23}$ cm$^{-2}$ and ionization state increasing
progressively from bottom to top with 
$\xi=0$ (black), 1
(cyan), 10 (red), 100 (green) and 1000 (blue) for photo-ionization by a power
law of $\Gamma=2$ as calculated by {\sc absori}. 
b) shows how the opacity of partially ionized
material ($\xi=100$) can instead be dominated by 
bound-bound (line) transitions using a proper photo-ionisation code 
such as {\sc xstar}.} 
\label{f:ionise} 
\end{figure*}

The ion populations are determined by the balance between
photo-ionization and electron recombination. For a given element $X$,
the ratio between $X^{+i}$ and the next ion stage up, $X^{i+1}$ is
given by the equilibrium reaction 
$$X^{+i} + h \nu \rightleftharpoons X^{+(i+1)} + e^-$$ The
photo-ionization rate depends on the number density of the ion
$N_X^{+i}$ and of the number density of photons, $n_\gamma$, above the
threshold energy $\nu_{edge}$ for ionization for that species. This
number density can be approximated by 
$n_\gamma \sim L /(h \nu 4 \pi r^2 c)$ where $L$ is
the source luminosity, $h \nu$ is the typical photon energy and $4 \pi
r^2 c $ is the volume swept out by the photons in 1 second. Actually,
what really matters is the number of photons past the 
threshold energy which photo-ionize the ion, so this depends on
spectral shape as well. 

Equilibrium
is where the photo-ionization rate, $N_X^{+i} n_{\gamma} \sigma(X^{+i})
c$ balances the recombination rate $N_X^{+(i+1)} n_e
\alpha(X^{+(i+1)}, T)$ where $\sigma(X^{+i})$ is the photoelectric
absorption cross-section for $X^{+i}$ and $\alpha(X^{+(i+1)}, T)$ is
the recombination coefficient for ion $X^{+(i+1)}$ at temperature $T$.
Hence the ratio of the abundance of the ion to the next stage down is 
given by
$$ {N_X^{+(i+1)} \over N_X^{+i}} = {n_{\gamma} \sigma(X^{+i}) c \over
n_e \alpha(X^{+(i+1)}, T)} \propto {n_\gamma\over n_e}$$ Thus the
ratio of photon density to electron density determines the ion
state. If the photon density is highest, the ion meets a photon first,
so is ionized to the next stage. Conversely, if the electron density
is higher, then the ion meets an electron first and recombines to the
lower ion stage. The ratio of photon to electron number density can be
written as $n_\gamma/n_e = \xi /(4 \pi h \nu c)$ where $\xi=L/nr^2$ is
the photo-ionization parameter. There are other ways to define this
such as $\Xi =P_{rad}/P_{gas}=L/(4 \pi r^2 c n_e k T) = \xi /(4 \pi c
k T)$ but whichever description is used, the higher the ionization
parameter, the higher the typical ionization state of each element.

In general, the equilibrium reaction means that there are at least two
fairly abundant ionization stages for each element, so as long as the
higher ionization stage is not completely ionized then there are
multiple edges from each element as each higher ion stage has a higher
edge energy as the electrons are more tightly bound. This can be
clearly seen in Fig.~\ref{f:ionise}a, where the H-like Oxygen
($O^{+7}$) edge at 0.87~keV is accompanied by the He-like ($O^{+6}$)
edge at 0.76~keV for $\xi=100$ and 10.

The {\sc absori} (Done et al 1992) model in {\sc xspec} calculates the
ion balance for a given (rather than self-consistently computed)
temperature and hence gives the photo-electric absorption opacity from
the edges. However, this can be very misleading as it neglects line
opacities (see below, and Fig.~\ref{f:ionise}b)

\subsection{Absorption Lines}
\label{s:lines}

\begin{figure*} 
\vskip -50pt
\begin{tabular}{l} 
\includegraphics[width=0.6\textwidth,angle=-90]{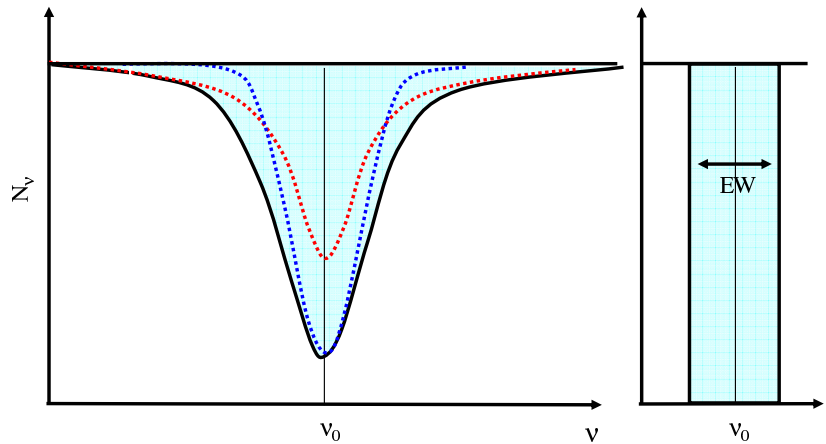} 
\end{tabular} 
\caption{a) The Voigt profile of an absorption line (black line), with
a Doppler core (dark grey/blue dotted line) and Lorentzian wings
(light grey/red dotted
line). b) The equivalent width of an absorption line is the width of a
rectangular notch in the spectrum at the rest wavelength which
contains the same number of photons as in the line.  }
\label{f:profile} 
\end{figure*}

There are also line (bound-bound) transitions as well as edges
(bound-free transitions).  These can occur whenever the higher shells
are not completely full. Hence Oxygen can show absorption at the n=1
to n=2 (1s to 2p) shell transition even in neutral material whereas
elements higher than Ne need to be ionized before this transition can
occur. The cross-section in the line depends on the line width. This
is described by a Voigt profile. The 'natural' line width is set by
the Heisenburg uncertainty relation between the lifetime of the
transition $\Delta t \Delta \nu \la \hbar $. This forms a Lorentzian
profile, with broad wings. However, the ions also have some velocity
due to the temperature of the material $v_{thermal}^2\sim
kT_{ion}/m_{ion}$. Any additional velocity structure such as
turbulence adds in quadrature so $v^2=v_{thermal}^2+v_{turb}^2$.
These velocities Doppler shift the transition, giving a Gaussian core
to the line.  This combination of Gaussian core, with Lorentzian
wings, is termed a Voigt profile (see Fig.~\ref{f:profile}a)

\begin{figure*} 
\vskip -50pt
\begin{tabular}{l} 
\includegraphics[width=0.7\textwidth,angle=-90]{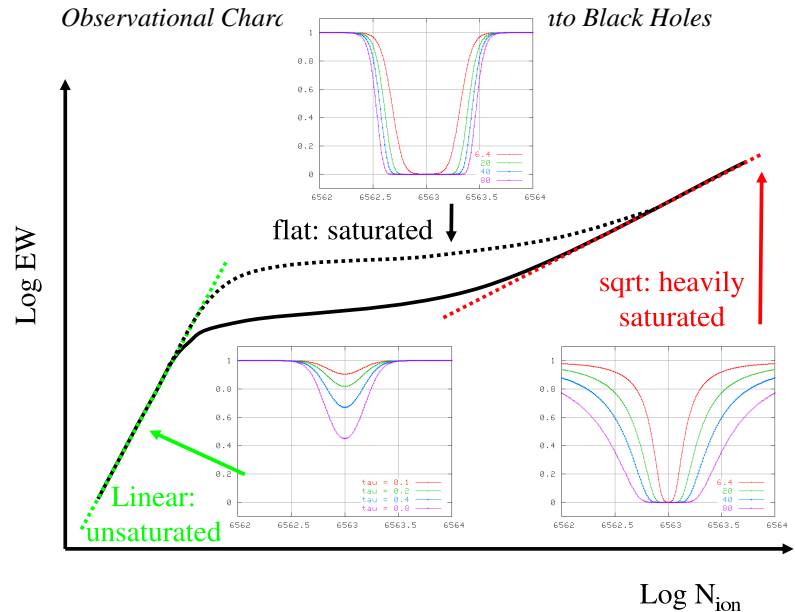} 
\end{tabular} 
\caption{A curve of growth. On the unsaturated section the line
equivalent width increases linearly with increasing column. However,
once the line becomes optically thick in the core, increasing the
column does not lead to an increase in absorption as the line center
is black. The Doppler wings are very steep so there is very little
change in equivalent width with column (saturated) until the line
becomes so optically thick that the Lorentzian wings dominate, leading
to the heavily saturated increase. The point at which the line becomes
saturated depends on the velocity structure. Higher turbulent velocity
means that the line is broader so it remains optically thin in the
core (unsaturated) up to a higher column density (dotted black line). 
}
\label{f:cog} 
\end{figure*}

The line equivalent width is the width of a rectangular notch (down to
zero) in the spectrum which contains the same number of photons as in
the line profile, as shown in Fig.~\ref{f:profile}b. The equivalent
width of the line grows linearly with the column density of the ion 
when the line is optically thin in its core, as the line gets deeper
linearly with column (called unsaturated). However, this linear
behaviour stops when the core of the line becomes optically thick i.e.
none of the photons at the line core can escape 
at the rest energy of the transition. Increasing the column cannot
lead to much more absorption as there are no more photons at the line
center to remove. The wings of the line can become optically thick but
Doppler wings are very steep so the line equivalent width does not
increase much as the column increases (called saturated).  Eventually, the
Lorentzian wings start to become important, and then the line
equivalent width increases as the square root of the column density
(called heavily saturated). This relation between column and line equivalent
width is termed a 'curve of growth'. While the linear section of this
is unique, the point at which the line becomes saturated depends on
the Doppler width of the line, i.e. on the velocity structure of the
material as shown in Fig.~\ref{f:cog}.

For 'reasonable' velocities, the line absorption equivalent width can
be larger than the equivalent width of the edges, so that line
transitions dominate the absorption spectrum. Fig~\ref{f:ionise}b
shows the total absorption (line plus edges computed using {\sc
warmabs}, a model based on the XSTAR photo-ionization code (Kallman \&
Bautista 2001; Kallman et al 2004) compared to the more approximate
ionization code which just used edge opacities ({\sc absori} in {\sc
xspec}). The differences are obvious. There are multiple ionized
absorption lines which dominate the spectrum as well as photo-electric
edges.

Soft X-ray absorption features, especially from H and He like Oxygen,
are seen in half of type 1 Seyferts (Reynolds 1997). High
resolution grating spectra from Chandra and XMM-Newton observations
have shown these in some detail. Typically, the best datasets require
several different column densities, each with different ionization
parameter, in order to fit the data.  By contrast, the BHB show little
in the way of soft X-ray absorption (GROJ1655-40 is an extreme
exception: Miller et al 2006a), but highly inclined systems do show H
and He--like Fe K$\alpha$ absorption lines in bright states (Ueda et
al. 1998; Yamaoka et al. 2001; Kotani et al. 2000; Lee et al. 2002;
Kubota et al 2007).

In each case, the origin of this ionized material can be constrained by
measuring the velocity width (or an upper limit on the velocity width)
of a line to get the column density of that ion via a curve of
growth. This can be done separately for each transition, but a better
way is to take a photo-ionization code and use this to calculate the
(range of) column density and ionization state required to produce all
the observed transitions assuming solar abundance ratios. The
spectrum gives an estimate for $L$ so the equation for the ionization
parameter $\xi=L/nR^2$ can be inverted by writing $N_H=n\Delta R$ to
give $R = L/(N_H \xi )\times (\Delta R/R)$. Since $\Delta R/R\la 1$,
then the distance of the material from the source is $R\la L/(N_H \xi
)$. A small radius means that the material is most probably launched
from the accretion disc itself, probably as a wind.

\subsection{Winds}
\label{s:winds}

Wherever X-rays illuminate material they can photo-ionize it. They
also interact with the electrons by Compton scattering. Electrons are
heated by Compton upscattering when they interact with photons of
energy $\epsilon \ga \Theta$ but are cooled by Compton scattering by
photons with energy $\epsilon \la \Theta$.  Since the illuminating
spectrum is a broadband continuum, then the spectrum both heats and
cools the electrons. The Compton temperature is the equilibrium
temperature where heating of the electrons by Compton downscattering
equals cooling by Compton upscattering.  Section \ref{s:comp} shows
that the mean energy shift is $\Delta \epsilon \sim 4 \Theta \epsilon -
\epsilon^2$, so integrating this over the photon spectrum gives the net
heating which is zero at the equilibrium Compton temperature,
$\Theta_{ic}$, so  $0=\int
N(\epsilon) \Delta \epsilon d\epsilon = \int N(\epsilon) (4
\Theta_{ic}\epsilon - \epsilon^2) d\epsilon$.  Hence
$$\Theta_{ic}= {\int N(\epsilon) \epsilon^2 d \epsilon \over 4 \int
N(\epsilon) \epsilon d\epsilon} $$ For a photon spectrum with $\Gamma
= 2.5$ between $\epsilon_i$ to $\epsilon_{max}$ this gives
$\Theta_{ic}\approx \frac{1}{4}\sqrt{\epsilon_{max}\epsilon_i}$ $\sim
2.5~keV$ for energies spanning 1-100~keV.  Alternatively, for a hard
spectrum with $\Gamma =1.5$ this is $\approx \epsilon_{max}/12$ or
8~keV.  The effective upper limit to $\epsilon_{max}$ is around
100~keV as the reduction in Klein-Nishina cross-section means that
higher energy photons do not interact very efficiently with the
electrons.

Thus the irradiated face of the material can be heated up to this
Compton temperature, giving typical velocities in the plasma of
$v^2_{ic}=3 k T_{ic}/m_p$. This is constant with distance from the
source as the Compton temperature depends only on spectral shape
(though the depth of the heated layer will decrease as illumination
becomes weaker).  This 
velocity is comparable to the escape velocity from the central object
when $v^2_{ic} \sim GM/R_{ic}$, defining a radius, $R_{ic}$, 
at which the Compton
heated material will escape as a wind (Begelman, McKee \& Shields
1983).  This is driven by the pressure gradient, so has typical
velocity at infinity of the sound speed $c_s^2 = kT_{ic}/m_p \sim
v^2_{ic} \sim GM/R_{ic}$. Thus the typical velocity of this thermally
driven wind is the escape velocity from where it was launched.

As the source approaches Eddington, the effective gravity is reduced
by a factor $(1- L/L_{Edd})$, so the thermal wind can be launched from 
progressively smaller radii as the continuum radiation pressure enhances
the outflow, forming a radiation driven wind. 

The Eddington limit assumes that the cross-section for interaction
between photons and electrons is only due to electron
scattering. However, where the material is not strongly ionized,
there are multiple UV transitions, both photo-electric absorption
edges and lines. This reduces the 'Eddington' limit by a factor
$\sigma_{abs}/\sigma_T$ which can be as large as 4000. This opacity is
mainly in the lines, and the outflowing wind has line transitions which
are progressively shifted from the rest energy to $\Delta \nu/\nu \approx
v_{\infty}/c$. This large velocity width to the line means that its
equivalent width is high, so it can absorb momentum from the line
transition over a wide range in energy. This gives a UV line driven
wind. 

The final way to power a wind is via magnetic driving, but this
difficult to constrain as it depends on the magnetic field
configuration, so it is invoked only as a last resort.

Much of the 'warm absorber' systems seen in AGN have typical
velocities, columns and ionization states which imply they are
launched from size scales typical of the molecular torus (e.g. Blustin
et al 2005). The much faster velocities of $\sim 0.1-0.2$~c implied by
the broad absorption lines (BAL's) seen bluewards of the corresponding
emission lines in the optical/UV spectra of some Quasars 
are probably a UV line driven wind from the
accretion disc.  However, the similarly fast but much more highly
ionized (H and He-like Fe) absorption systems seen in the X-ray
spectra of some AGN (see ahead in Fig~\ref{f:mkn766}c and d) 
probably require either
continuum driving with $L \sim L_{Edd}$ or magnetic driving (as the
ionization state is so high that the line opacity is negligible with
respect to $\sigma_T$).

By contrast, the BHB typically show fairly low outflow velocity of the
highly ionized Fe K$\alpha$, mostly consistent with a thermally driven
wind from the outer accretion disc (e.g. Kubota et al 2007, DGK07), although the extreme absorber seen in one observation from
GRO~J1655-40 may require magnetic driving (Miller et al 2006a, 
Kallman et al 2009). However, this
does assume that the observed luminosity, $L_{obs}$ measures the
intrinsic luminosity, $L_{int}$. If there is electron scattering in
optically thick, completely ionized material along the line of sight
then $L_{obs}=e^{-\tau} L_{int} \ll L_{int}$ (Done \& Davies
2008). Such scattering would strongly suppress the rapid variability
power (Zdziarski et al 2010) and indeed 
the variability power spectra of these data lack all high
frequency power above 0.3-1~Hz, rather than extending to the
$\sim 10$~Hz seen normally
(see Fig.~\ref{f:states}). Thus thermal winds potentially explain 
all of what we see in terms of absorption from BHB. 

\section{Reflection}
\label{s:refl}

\begin{figure*}
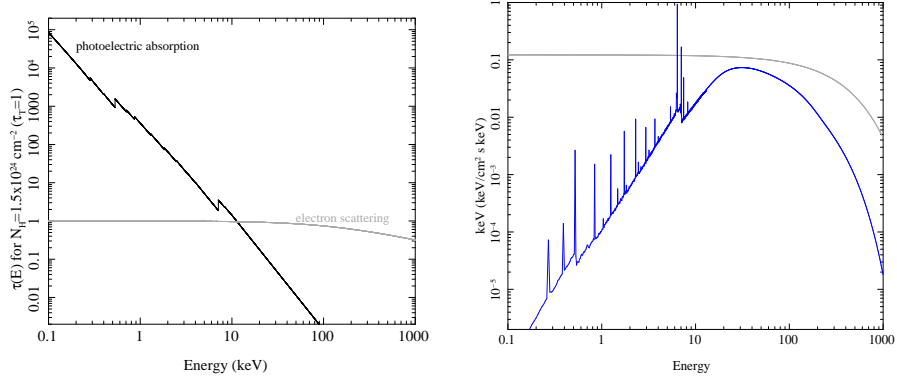
 
\begin{tabular}{cc} 
\includegraphics[width=0.45\textwidth]{cross-x_ne.ps} 
&
\includegraphics[width=0.45\textwidth,height=0.4\textwidth]{refl_col.ps}
\end{tabular} 
\caption{a) The absorption cross-section for neutral material with
solar abundance for a column of $\tau_T=1$ i.e. $N_H=1.5\times
10^{24}$ cm$^{-2}$ (black), together with the full (Klein-Nishima) electron
scattering cross-section (grey). b) shows the corresponding reflection
spectrum from such material. The much larger absorption cross-section
at low energies means that most incident photons are absorbed rather
than reflected, but Compton downscattering means that high energy
photons are not reflected elastically. The combination of these two
effects makes the characteristic peak between 20-50~keV, termed the
reflection hump, with atomic
features superimposed on this. Figure courtesy of M. Gilfanov.}
\label{f:refl} 
\end{figure*}

Whereever X-rays illuminate optically thick material such as the
accretion disc the photons have some probability to scatter off an
electron, and so bounce back into the line of sight. This reflection
probability is set by the relative importance of scattering versus
photo-electric absorption.  For neutral material, photo-electric
absorption dominates at low energies so the reflected fraction is very
small. However, the photo-electric cross-section decreases with energy
so the reflected fraction increases. Iron is the last astrophysically
abundant element (due to element synthesis in stars as released in
supernovae - see P. Podsiadlowski, this volume), so after 7.1 keV
there are no more significant additional sources of opacity. The
cross-section decreases as $E^{-3}$, becoming equal to $\sigma_T$ at
around 10~keV for solar abundance material (Fig.~\ref{f:refl}a).
Above this, scattering dominates, leading to a more constant reflected
fraction, but at higher energies, the photon energy is such that
Compton downscattering is important, so the reflection is no longer
elastic. Photons at high energy do reflect, but do not emerge at the
same energy as they are incident. This gives a break at high energies
as Compton scattering conserves photon {\em number}, and the number of
photons is much less at higher energies. Thus neutral reflection gives
rise to a very characteristic peak between 20-50~keV, termed the
reflection hump, where lower energy photons are photo-electrically
absorbed and higher photons are (predominantly) downscattered
(Fig.~\ref{f:refl}b: George \& Fabian 1991; Matt, Perola \& Piro 1991)

The dependence on photo-electric absorption at low energies means that
the spectrum is also accompanied by the associated emission lines as
the excited ion with a gap in the K (n=1) shell decays to its ground
state. This excess energy can be emitted as a fluorescence line
(K$\alpha$ if it is the n=2 to n=1 transition, K$\beta$ for n=3 to n=1
etc). However, at low energies, the reflected emission forms only a
very small contribution to the total spectrum, so any emission lines
emitted below a few keV are strongly diluted by the incident
continuum. These lines are also additionally suppressed as low atomic
number elements have a higher probability to de-excite via Auger
ionization, ejecting an outer electron rather than emitting the excess
energy as a fluorescence line. This means that iron is the element
which has most impact on the observed emission, as this is emitted
where the fraction of reflected to incident spectrum is large, and has
the highest fluorescence probability.  All this combines to make a
reflection spectrum which contains the imprint of the iron K edge and
line features, as well as the characteristic continuum peak between
20-50~keV (Fig.~\ref{f:refl}b: George \& Fabian 1991; Matt, Perola \&
Piro 1991).

\subsection{Ionised Reflection}

\begin{figure*}
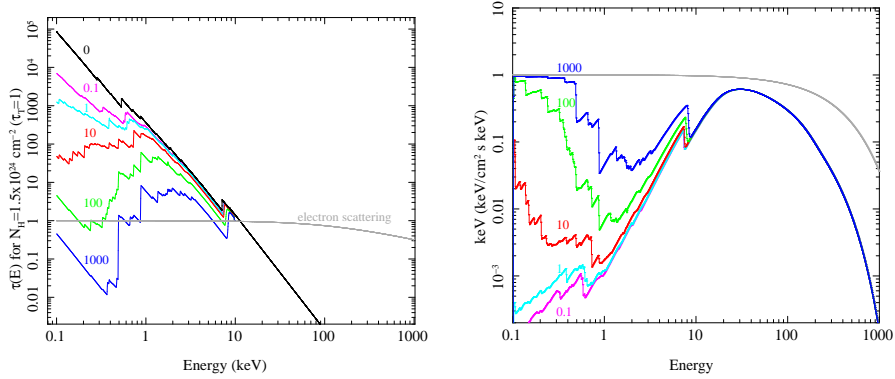
 
\begin{tabular}{cc} 
\includegraphics[width=0.45\textwidth]{cross-x.ps} 
&
\includegraphics[width=0.45\textwidth,height=0.4\textwidth]{pexriv.ps}
\end{tabular} 
\caption{a) The ionized absorption cross-section for material with
solar abundance for a column of $\tau_T=1$ i.e. $N_H=1.5\times
10^{24}$ cm$^{-2}$. The curves are labeled with the value of $\xi$, 
and shown together with the full (Klein-Nishima) electron
scattering cross-section (grey) as in Fig.~\ref{f:refl} 
b) shows the corresponding reflection
spectrum from such material. Increased ionization gives decreased
opacity at low energies and hence more reflection. The reflected
spectrum does not depend on photo-electric absorption at high
energies, so is unchanged by ionization.}
\label{f:ionref} 
\end{figure*}

The dependence on photo-electric absorption at low energies means that
reflection is sensitive to the ionization state of the reflecting
material.  Fig.~\ref{f:ionref}a shows how the absorption cross-section
changes as a function of ionization state using a very simple model
for photoelectric absorption which considers only the photoelectric
edge opacity. The progressive decrease in opacity at low energies for
increased ionization state means an increase in reflectivity at these
energies as shown in Fig.~\ref{f:ionref}b for simple models of the
reflected continuum ({\sc pexriv} model in {\sc xspec}). 

\begin{figure*}
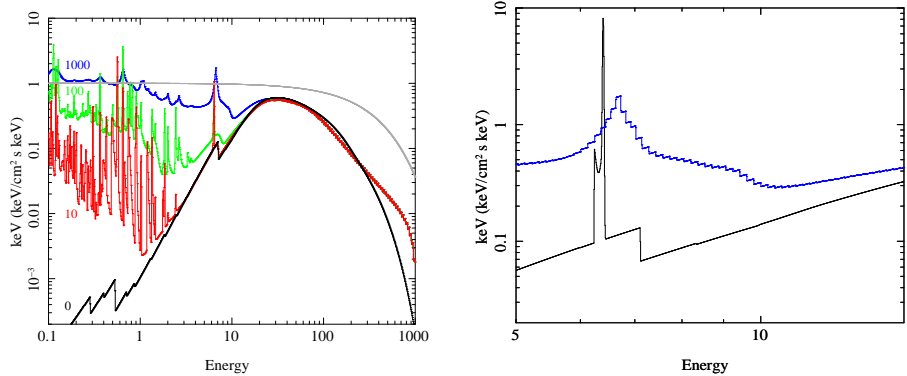
 
\begin{tabular}{cc} 
\includegraphics[width=0.45\textwidth]{reflion.ps} 
&
\includegraphics[width=0.45\textwidth,height=0.4\textwidth]{line.ps}
\end{tabular} 
\caption{a) Ionized reflection models from a constant density slab
which include the self-consistent line and recombination continuum
emission.  The curves are labeled with the value of $\xi$, and a
clearly much more complex than the simple reflected continua models
in Fig.~\ref{f:ionref}. b) shows a detailed view of the iron line
region. For neutral material (black), around 1/3 of the line photons
are scattered in the cool upper layers of the disc before escaping,
forming a Compton downscattered shoulder to the line. Conversely, for
highly ionised reflection (blue), the upper layers of the disc are
heated to the Compton temperature and Compton upscattering as well as
downscattering broadens the spectral features.}
\label{f:reflion} 
\end{figure*}

However, this continuum is accompanied by emission lines and
bound-free (recombination) continua and the lines can be more
important for ionized material. This is due to both an increase in
emissivity (He-like lines especially have a high oscillator strength)
and to the fact that the increased reflected fraction at low energies
mean that these lines are not so diluted by the incident
continuum. Better models of ionized reflection are shown in
Fig.~\ref{f:reflion}. These include both the self consistent emission
lines and recombination continua, and the effects of Compton
scattering within the disc.  By definition, we only see down to a
depth of $\tau(E)=1$. Thus the reflected continuum only escapes from
above a depth of $\tau(E) \sim 1$. Fig.~\ref{f:refl}a shows that the
iron line is produced in a region with $\tau_T\sim 0.5$, so a fraction
$e^{-\tau_T} \sim 1/3$ of the line is scattered. For neutral
material, this forms a Compton downscattering
shoulder to the line, but for ionized material the disc is
heated by the strong irradiation up to the Compton
temperature. Compton upscattering can be important as well as
downscattering, so the line and edge features are broadened
(Young et al 2001, see Fig.~\ref{f:reflion}b). Models including
these effects are publically available as the {\sc xspec} 
{\sc atable} model, {\sc reflionx.mod}, and this should be used rather
than {\sc pexriv} for ionized reflection. However, the incident continuum
for this model is an exponential power law, so this can have problems
with the continuum form at high energies (see Fig. \ref{f:xspec_comp}a).

\subsection{Ionization Instability: Vertical Structure of the disc}
\label{s:instability}

\begin{figure*} 
\vskip -50pt
\begin{tabular}{l} 
\includegraphics[width=0.7\textwidth,angle=-90]{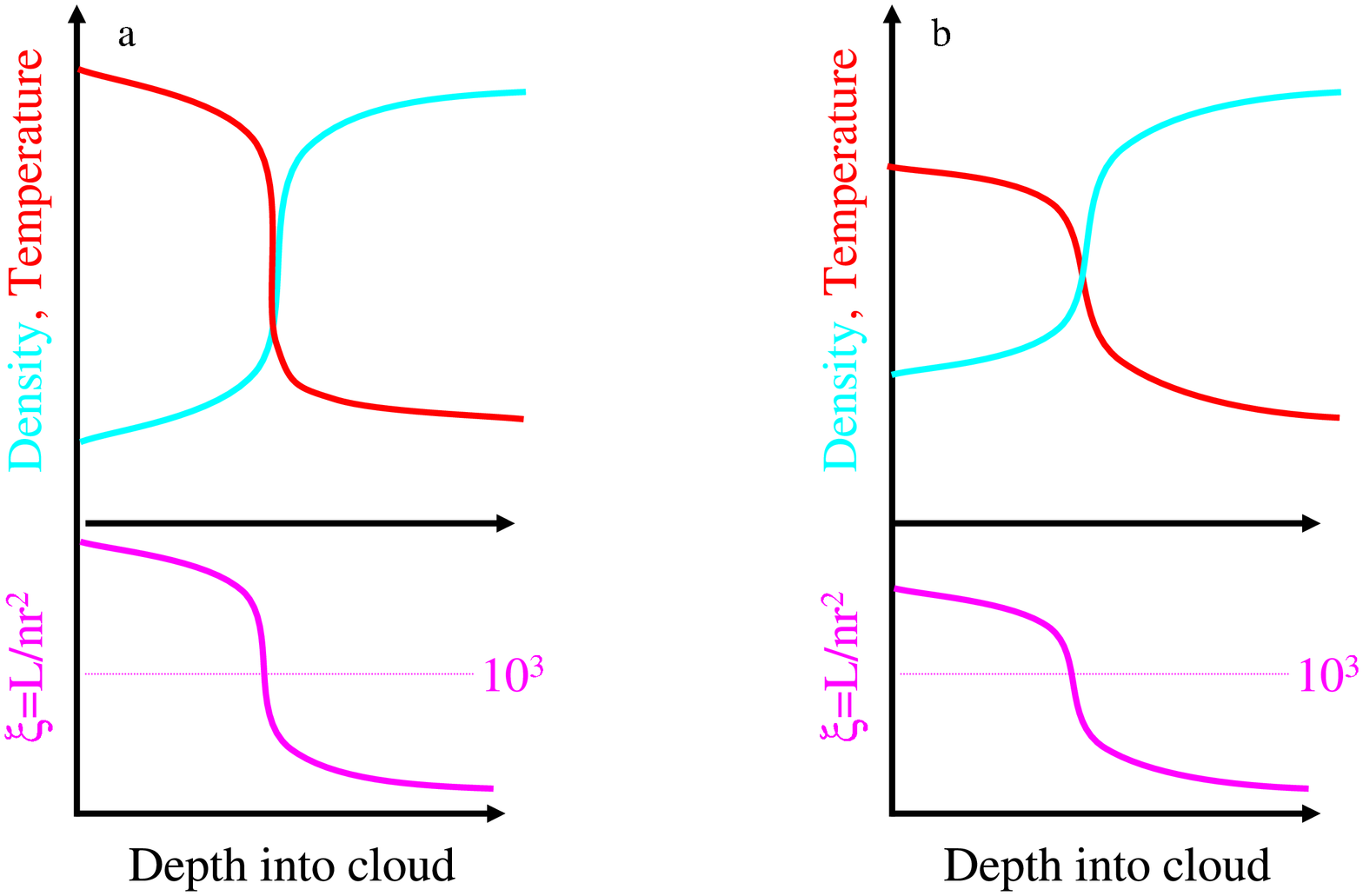} 
\end{tabular} 
\caption{a) The ionization instability for material in hydrostatic
equilibrium (or any general pressure balance) from 
hard spectral
illumination. The high Compton temperature means that the material
expands to very low density, and hence high ionization. b) shows the
corresponding vertical structure from soft spectral illumination,
where the lower Compton temperature gives higher surface density and
hence lower ionization. Both show the rapid drop in
ionization that comes from the dramatic increase in cooling from lines
when partially ionized ions can exist.}
\label{f:instability} 
\end{figure*}

\begin{figure*} 
\begin{tabular}{cc} 
\includegraphics[width=0.45\textwidth]{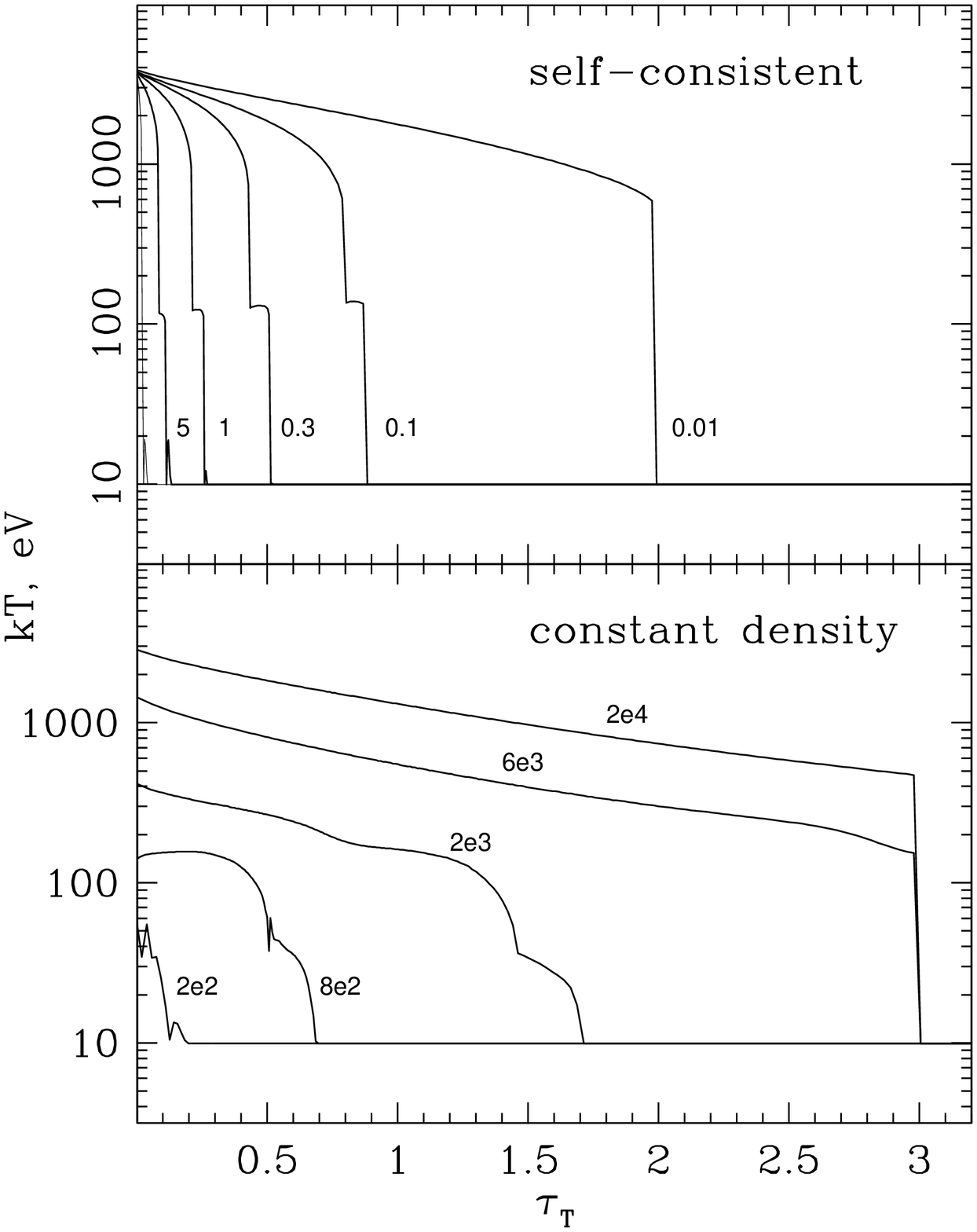} 
&
\includegraphics[width=0.45\textwidth]{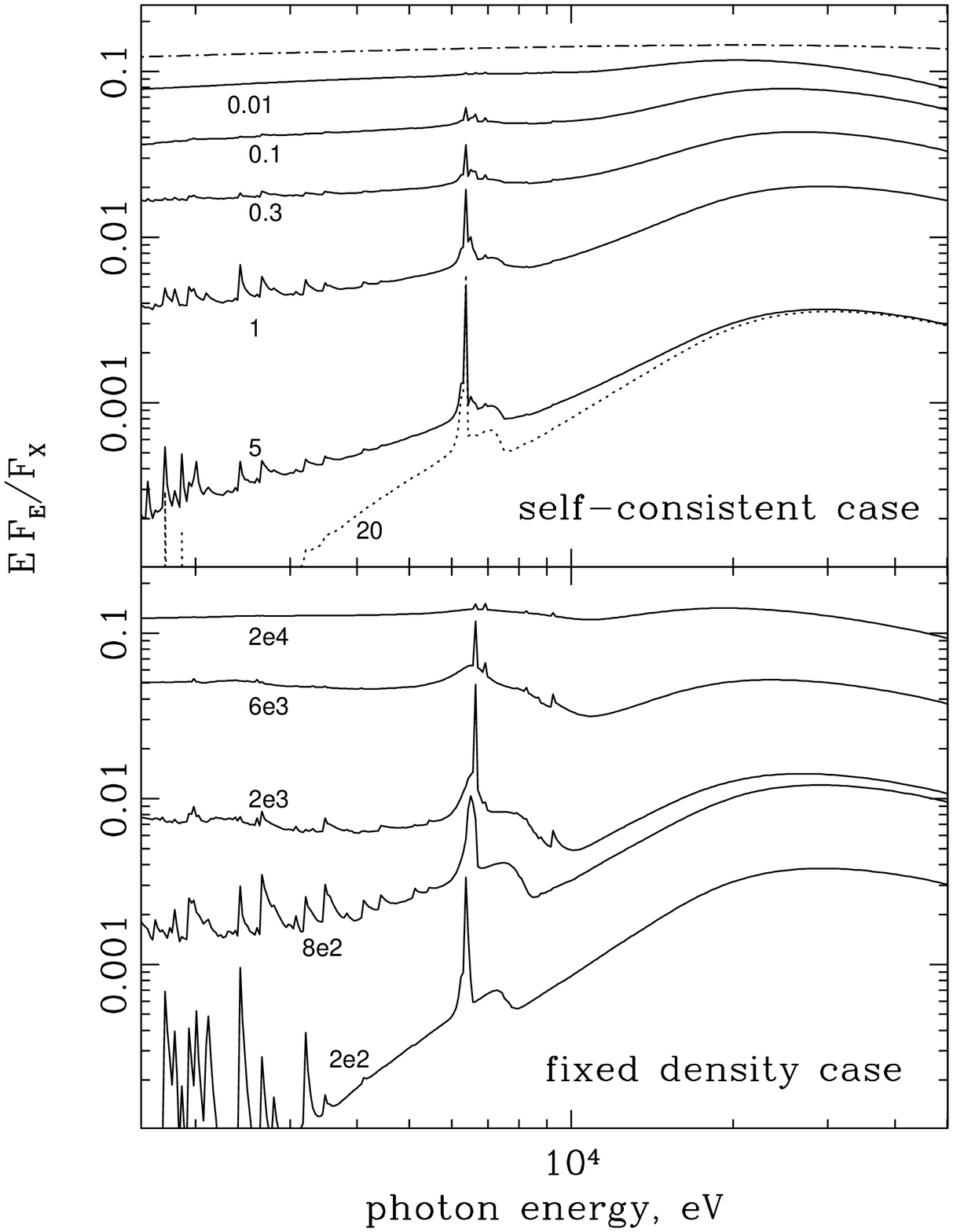} 
\end{tabular} 
\caption{a) The vertical temperature structure from material in
hydrostatic equilibrium (upper panel) shows a sharp drop in
temperature due to the ionization instability. This is rather
different to the much smoother drop in temperature seen for assuming
the material has constant density (lower panel). b) The resulting
differences in ionization structure of the disc photosphere give rise
to different reflected spectra. In particular, the deep edge seen in
the constant density reflected spectrum at $\xi=2000$ is never present
in the hydrostatic models (Nayakshin et al 2000).}
\label{f:xion} 
\end{figure*}

The slab models described above assume that the irradiated material
has constant density. Yet if this material is a disc, then it should
be in hydrostatic equilibrium, so the density responds to the
irradiation heating. The top of the disc is heated to the Compton
temperature, so expands, so its density drops, so its ionization state
is very high. Further into the disc, hydrostatic equilibrium means that
the pressure has to increase in order to hold up the weight of the
layers above. The Compton temperature remains the same, so the density
has to increase, but an increase in density means an increase in
importance of bremsstrahlung cooling.  This pulls the temperature down
further, but the pressure must increase so the density has to increase
further still, so the cooling increases. Eventually the
temperature/ionization state drops to low enough levels that not all
the material is completely ionized. Bound electrons means that line
cooling can contribute, pulling the temperature down even faster, with
a corresponding increase in density. The material thus makes a very
rapid transition from almost completely ionized to almost completely
neutral. Thus the reflection spectrum is a composite of many different
ionization parameters, with some contribution from the almost
completely ionized skin, and some from the almost completely neutral
material underlying the instability point, but with very little
reflection at intermediate ionization states (Nayakshin, Kazanas \&
Kallman 2000; Done \& Nayakshin 2007). 

The difference in Compton temperature for hard spectral illumination
and soft spectral illumination only changes the ionization state of
the skin. For hard illumination, the high Compton temperature gives a
very low density and high ionization state so the skin is almost
completely ionized as described above. For softer illumination, the
Compton temperature is lower, so the density is higher and the
ionization state of the skin is lower, making it highly ionised rather
than completely ionized. However, there is still the very rapid
transition from highly ionized to almost neutral due to the extremely
rapid increase in cooling from partially ionized material
(Fig.~\ref{f:instability}; see also Done \& Nayakshin 2007).
Fig. ~\ref{f:xion}a and b shows how this very different
vertical structure for temperature-density-ionization 
affects the expected reflection signature.

The same underlying ionization instability, but for X-ray illumination
of dense, cool clouds in pressure balance with a hotter, more diffuse
medium (Krolik, McKee \& Tarter 1981), may well be the origin of the
multiple phases of ionization state seen in the AGN 'warm absorbers'
(e.g. Netzer et al 2003; Chevallier et al 2006).

\subsection{Radial Structure}

This vertical structure of a disc should change with radius, giving
rise to a different characteristic depth of the transition and hence a
different balance between highly ionized reflection from the skin and
neutral reflection from the underlying material. This does depend on
the unknown source geometry as well as the initial density structure
of the disc, and how it depends on radius, which in turn depends on
the (poorly understood) energy release in the disc (see e.g. Nayakshin
\& Kallman 2001). The {\sc xion} model (Nayakshin)
incorporates both the vertical structure from the ionisation instability,
and its radial dependence for some assumed X-ray geometry and
underlying disc properties. 

However, if the material is mostly neutral, then neither vertical nor
radial structure gives rise to a change in the reflected spectrum with
radius. Neutral material remains neutral as the illumination gets
weaker, so these models are very robust. 

\subsection{Relativistic Broadening}
\label{s:rel}

The reflected emission from each radius has to propagate to the
observer but it is emitted from material which is rapidly rotating in
a strong gravitational field. There is a combination of effects which
broaden the spectrum.  Firstly, the line of sight velocity gives a
different Doppler shift from each azimuth, with maximum blueshift from
the tangent point of the disc coming towards the observer, and
maximum redshift from the tangent point on the receding side. Length
contraction along the direction of motion means that the emission is
beamed forward, so the blueshifted material is also brightened while
the redshifted side of the disc is suppressed. These effects are
determined only by the component of the velocity in the line of sight,
so are not important for face on discs. However, the material is
intrinsically moving fast, in Keplerian rotation, so there is always
time dilation (fast clocks run slow, also sometimes termed
transverse redshift as it occurs even if the velocity is completely
transverse) and gravitational redshift.

All these effects decrease with increasing radius. The smaller
Keplerian velocity means smaller Doppler shifts and lower boosting
giving less difference between the red and blue sides of the line. The
lower velocity also means less time dilation while the larger radius
means less gravitational redshift. Thus larger radii give narrower
lines, so the line profile is the inverse of the radial profile, with
material furthest out giving the core of the line and material at the
innermost orbit giving the outermost wings of the line
(Fabian et al 1989; Fabian et al 2000). The relative weighting between the
inner and outer parts of the line are given by the radial emissivity,
where the line strength $\propto r^{-\beta}$. This gives 
$\beta=3$ for either an emissivity which follows the illumination pattern
from a gravitationally powered corona, or 'lamppost' point source
illumination. This characteristic
line profile is given by the {\sc diskline} (Fabian et al 1989)
and {\sc laor} (Laor 1991) models for Schwarzschild and extreme Kerr
spacetimes, respectively. 

\begin{figure*}
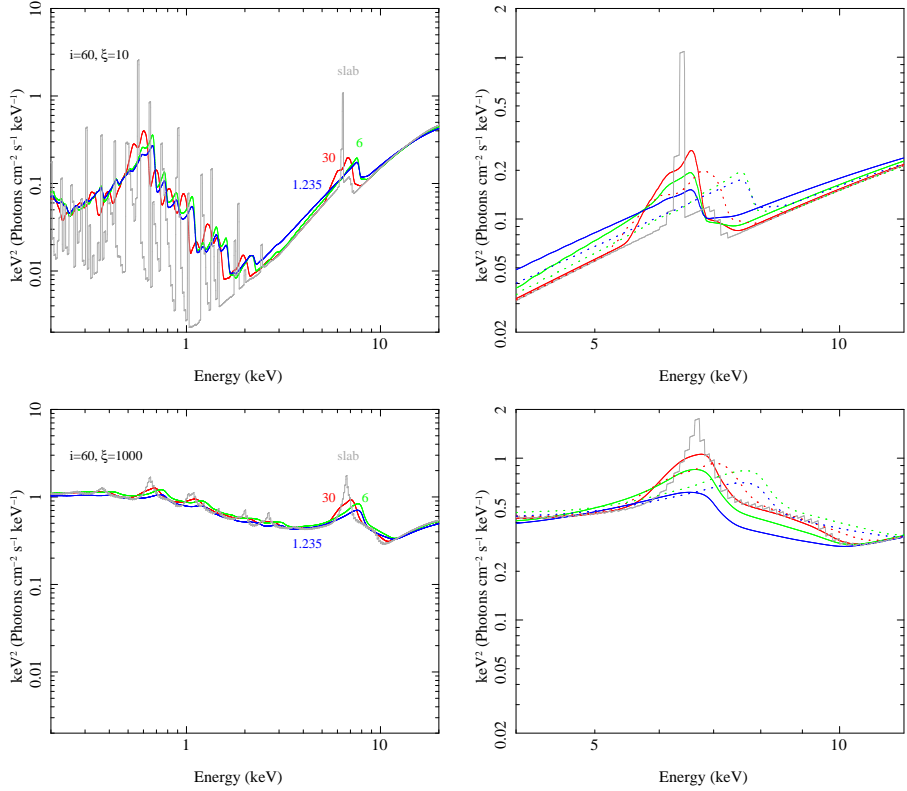
 
\begin{tabular}{cc} 
\includegraphics[width=0.45\textwidth]{xi10_i60.ps} 
&
\includegraphics[width=0.45\textwidth]{xi10_i30_i60_iron.ps}\\
\includegraphics[width=0.45\textwidth]{xi1000_i60.ps}
&
\includegraphics[width=0.45\textwidth]{xi1000_i30_i60_iron.ps}
\end{tabular} 
\caption{a) Relativistic smearing of reflection from a constant
density slab with low ionization ($\xi=10$) viewed at $60^\circ$. The
intrinsic emission (grey) is convolved with the relativistic functions
with $r_{in}=30$ (red), $6$ (green) and 1.235 (blue). b) focuses in on
the iron line region to show the difference in line profile for
$i=30^\circ$ (solid lines) compared to $i=60^\circ$ (dotted lines). 
c) and d) show the same for highly ionized material ($\xi=1000$). 
The relativistic effects shift the shape of the 
become much less marked as the spectral
features are intrinsically broadened by Comptonization in the strongly
irradiated material.}
\label{f:rel} 
\end{figure*}

However, these relativistic effects 
should be applied to the entire reflected continuum, not just the line.
This can be modeled with 
the {\sc kdblur} model (a re-coding of the {\sc laor} model for convolution), 
or the newer
{\sc ky} models which work for any spin (Dovciak, Karas \& Yaqoob 2004). 
Fig.~\ref{f:rel}a shows the {\sc reflionx.mod} reflection {\sc atable} 
convolved with {\sc kdblur} for
$r_{in}=30, 6$ and $1.23 R_g$ for $i=60^\circ$. Blueshifts slightly
predominate over redshifts, with the 'edge' energy (actually
predominantly set by the blue wing of the line) at 7.8~keV for $r_{in}$
=6 and 1.23 (green and blue, respectively) compared to 7.1~keV for
$r_{in}$ = 30 (red) and in the intrinsic slab spectrum
(grey). Redshifts are more important for lower
inclinations. Fig.~\ref{f:rel}b shows a comparison of the iron line
region for $i=60^\circ$ (dotted lines) with $i=30^\circ$ (solid
lines). The 'edge' energy is now $\sim 6.7$~keV for both $r_{in}$
=1.235 and 6.

Relativistic smearing is harder to disentangle for ionized material as
the iron features are broadened by Compton scattering so there are no
intrinsically sharp features to track the relativistic effects (see
Fig.~\ref{f:rel}c and d). Nonetheless, the energies at which the
line/edge features are seen are clearly shifted. 

\subsection{Observations of reflected emission in AGN: iron line and    
soft X-ray excess}

Reflection is seen in AGN. There is a narrow, neutral iron line
and reflection continuum from illumination of the torus, and there is
also a broad component from the disc. This broad component is often
consistent with neutral reflection from material within $50 R_g$
produced from $r^{-3}$ illumination (Nandra et al 2007). However, a
small but significant fraction of objects require much more extreme
line parameters, with $r_{in}< 3 R_g$, and much more centrally
concentrated illumination $\propto r^{-\beta}$ where
$\beta=5-6$!!(MCG-6-30-15: Wilms et al 2001; Fabian \& Vaughan 2003;
1H0707-495: Fabian et al 2002; Fabian et al 2004; NGC4051: Ponti et al
2006). These all generally high mass accretion rate objects,
predominantly Narrow Line Seyfert 1s.

\begin{figure*}
\begin{tabular}{cc}
\includegraphics[width=0.45\textwidth,clip=true]{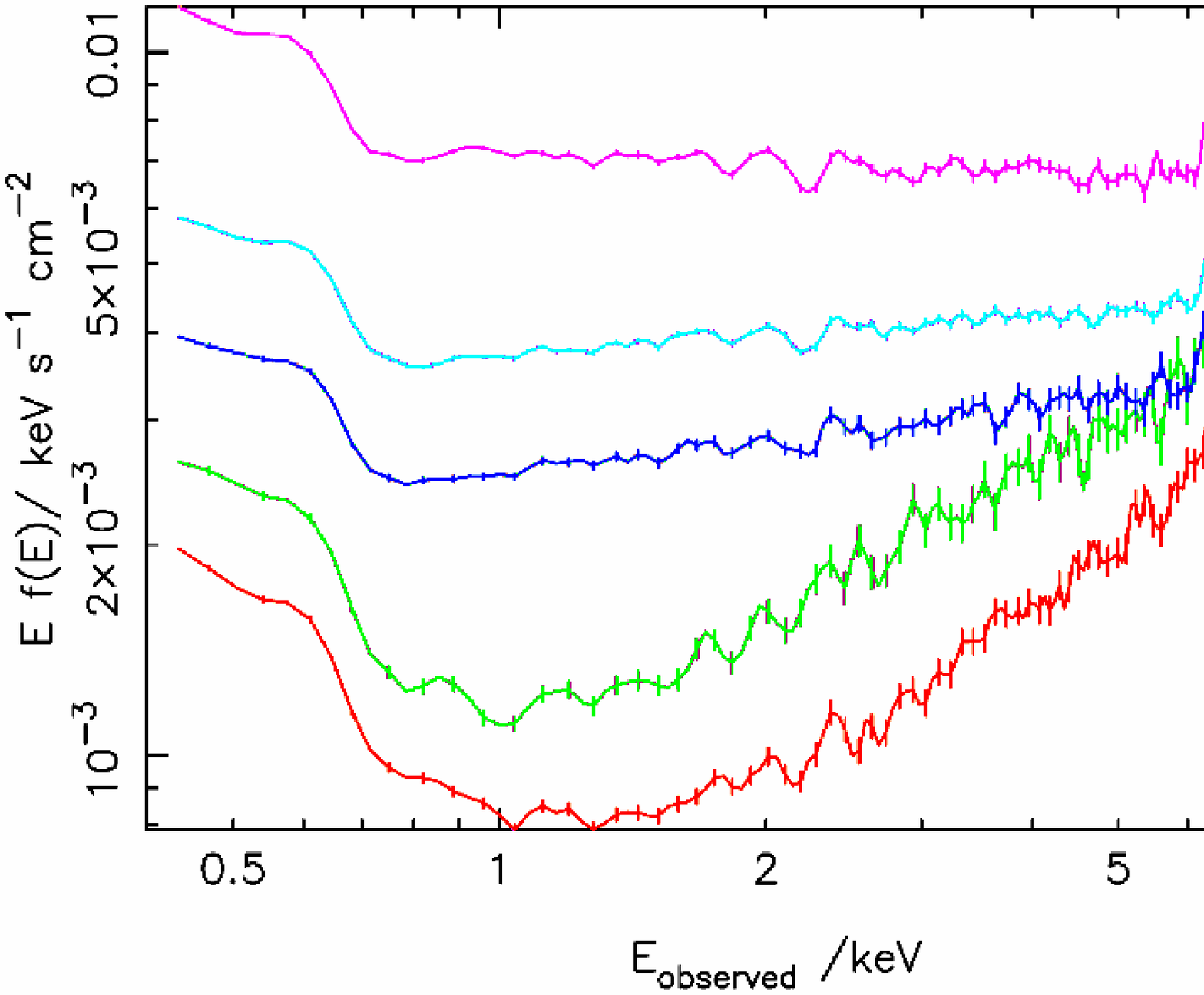}
&
\includegraphics[width=0.45\textwidth,clip=true]{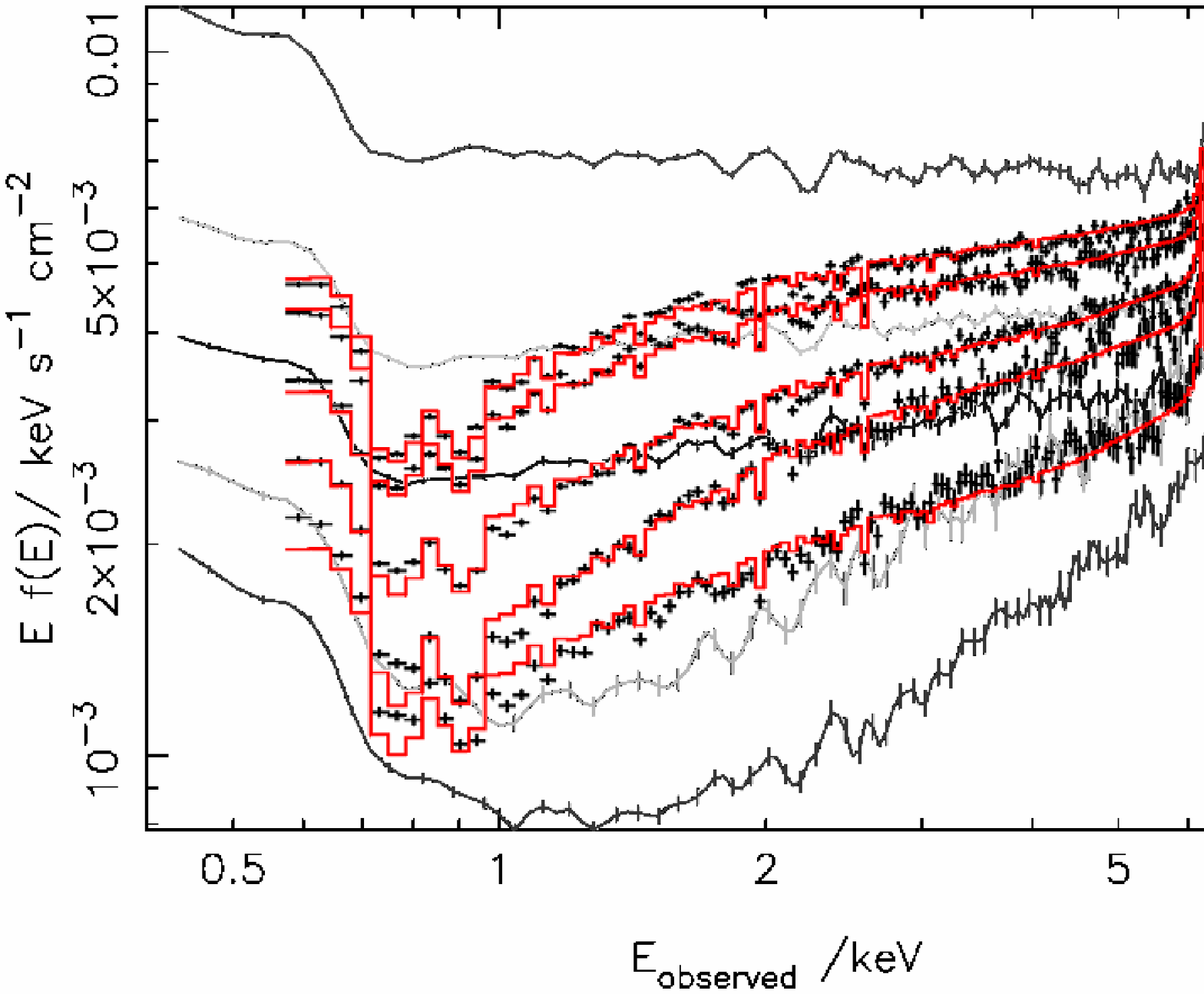}\\
\includegraphics[width=0.45\textwidth,clip=true]{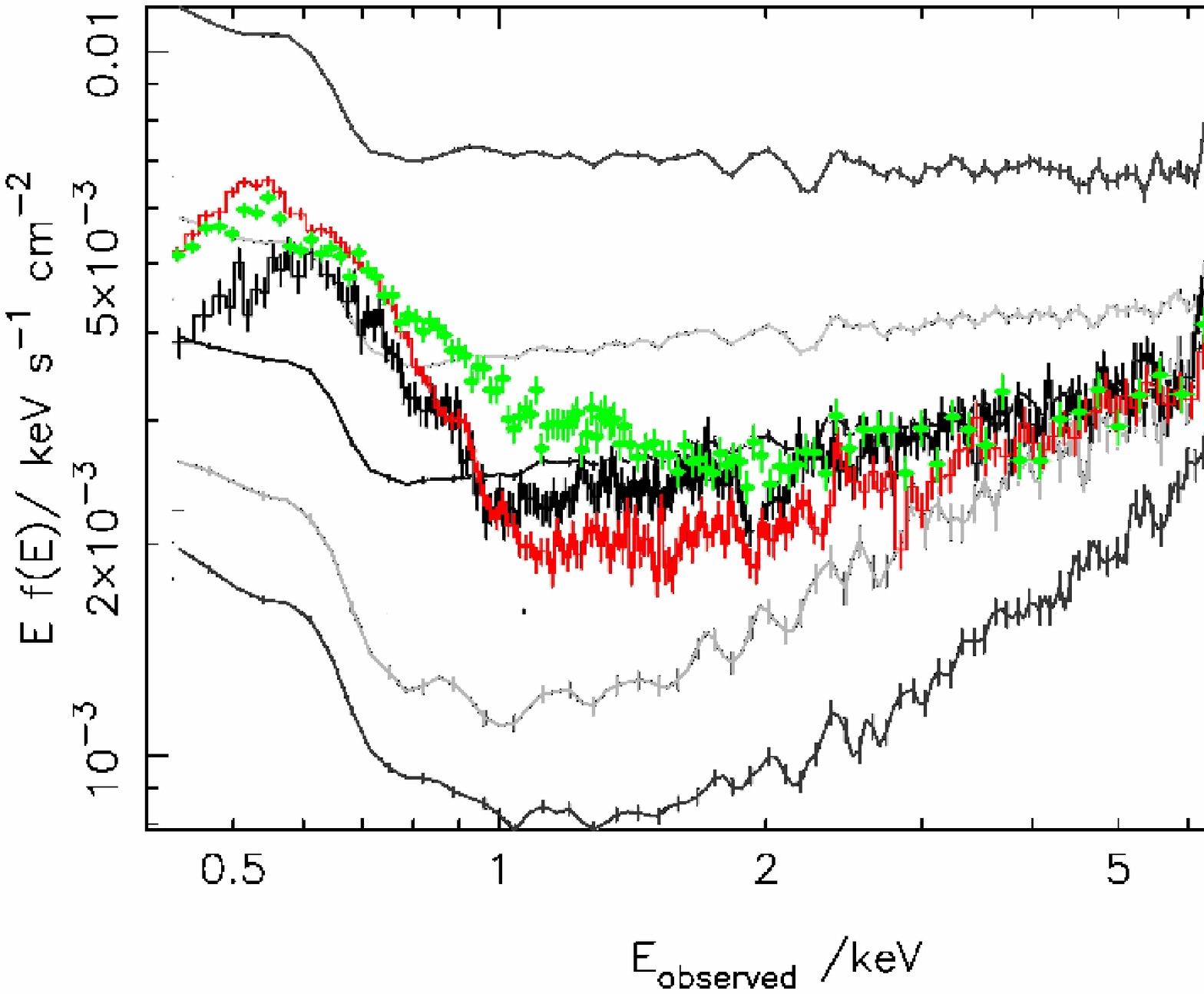}
&
\includegraphics[width=0.45\textwidth,clip=true]{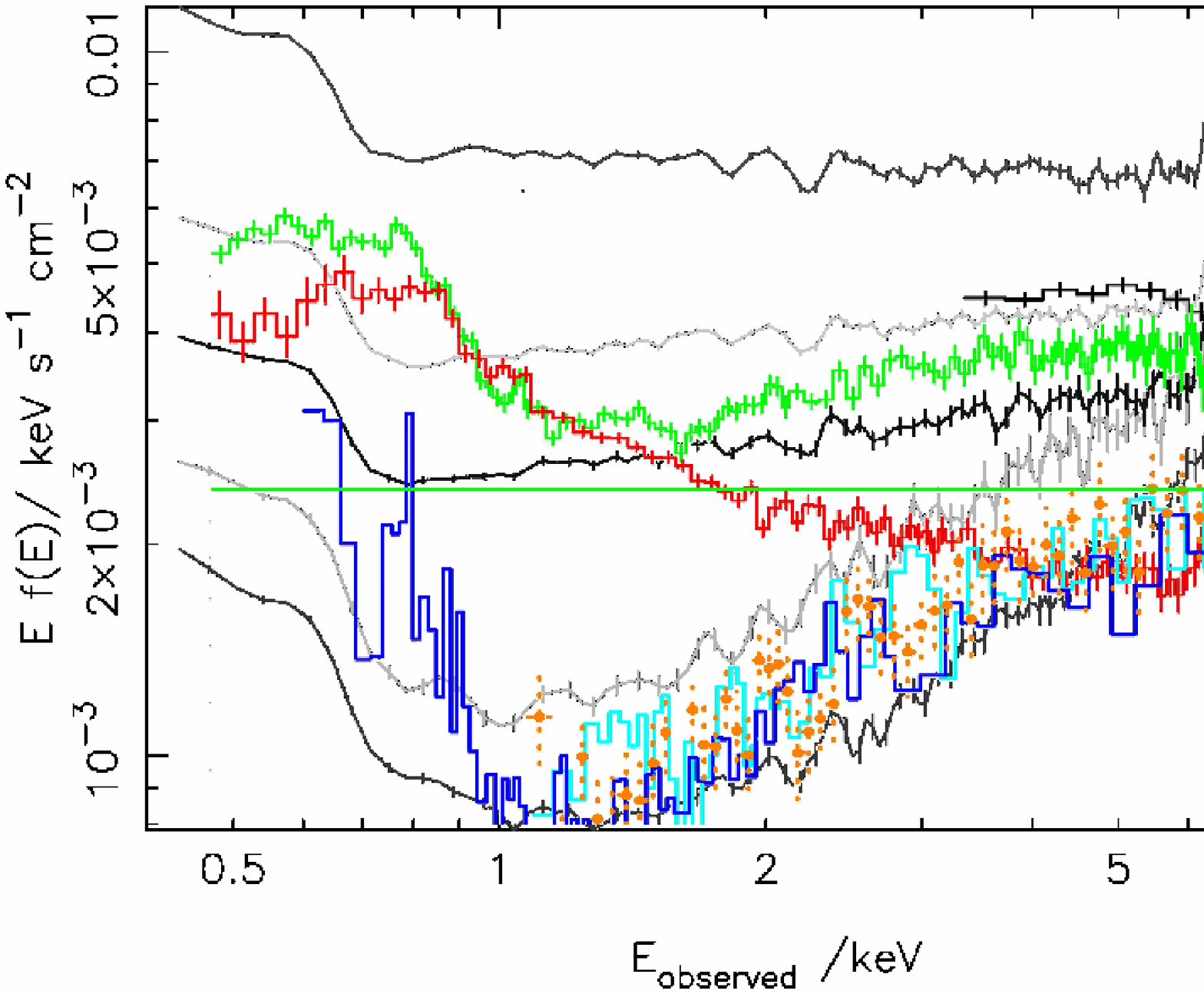}
\end{tabular}
\caption{Intensity sorted spectra from a selection of high mass
accretion rate AGN. a) shows 
Mkn766 (L. Miller et al 2007). This has a weak soft excess and soft 2--10 keV 
spectrum, together with a weak emission line at the highest luminosities,
while the lowest flux states have a strong soft excess, hard 2-7~keV spectrum,
with a strong drop above this, enhanced by strong He- and H-like absorption lines.
b) shows the same for MCG-6-30-15 
(L. Miller et al 2008). This shows very similar variability, though spans a smaller
range. It also has a stronger 'warm absorber', seen by the drop at 0.7-1~keV,
and stronger absorption lines of He and H--like iron on the blue wing of the line.
c) shows PG1211+10, with even less range in variability but with
ionized iron absorption which is strongly blueshifted (Reeves et al 2008; Pounds
et al 2003) d) shows PDS456 (Reeves et al 2009), 
where the lowest intensity spectra look like those
from MKN766, but with a stronger drop blueward of the line and even
more highly blueshifted iron absorption lines. However, this can also show
a rather different spectral shape, with steeper continuua (red points). 
}
\label{f:mkn766}
\end{figure*}

\begin{figure*}
\begin{tabular}{ccc}
\includegraphics[width=0.3\textwidth,clip=true]{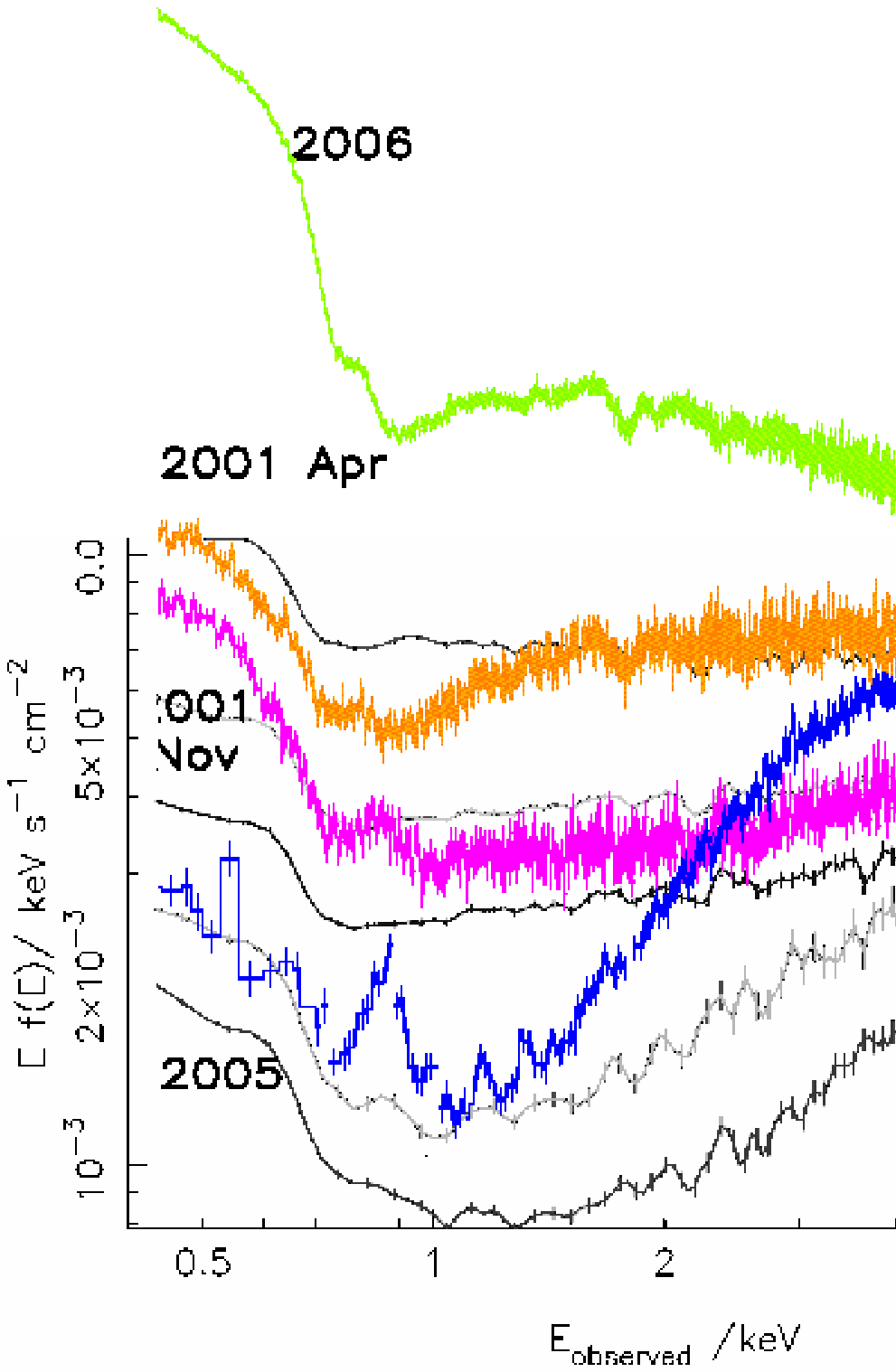}
&
\includegraphics[width=0.3\textwidth,clip=true]{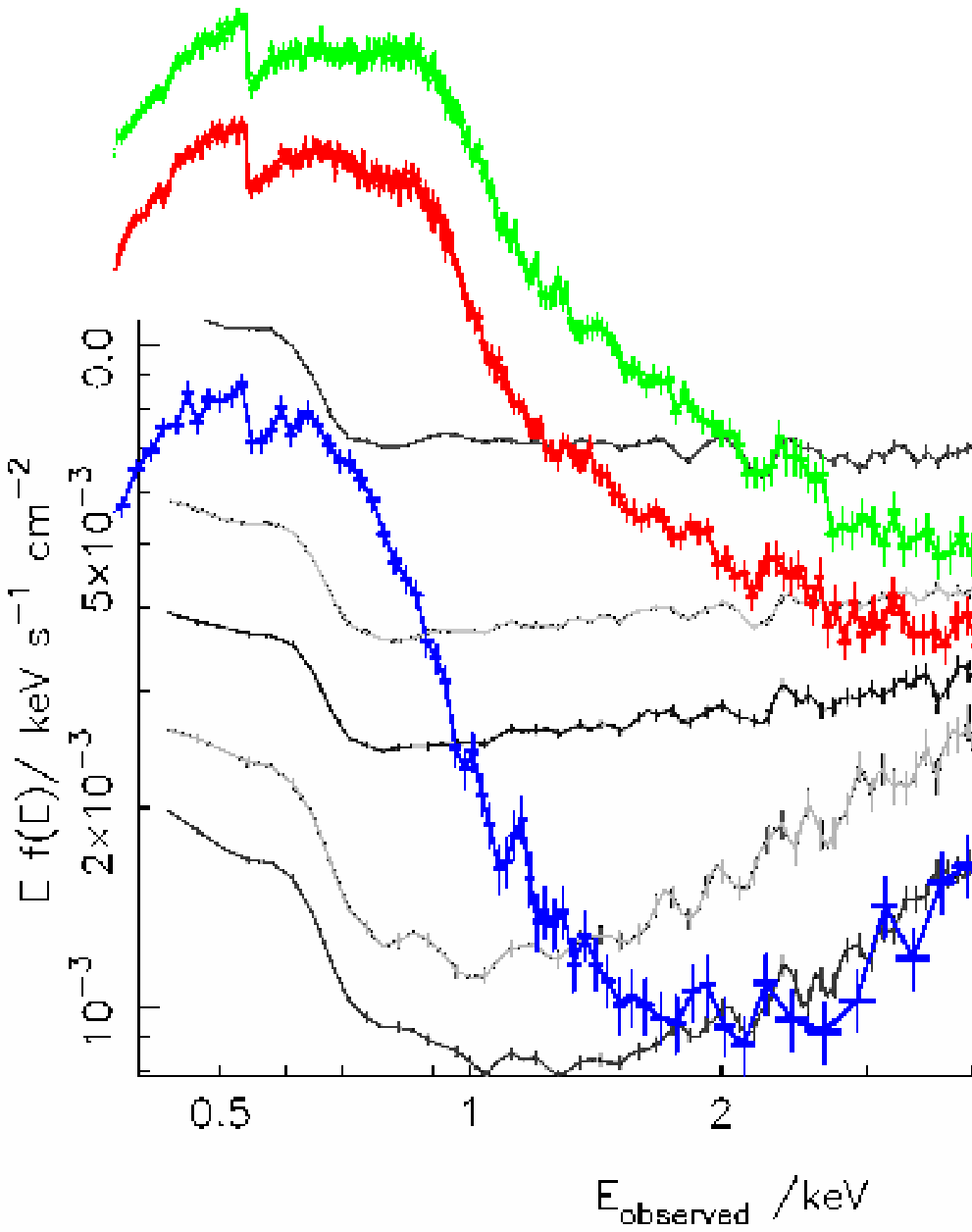}

&
\includegraphics[width=0.3\textwidth,clip=true]{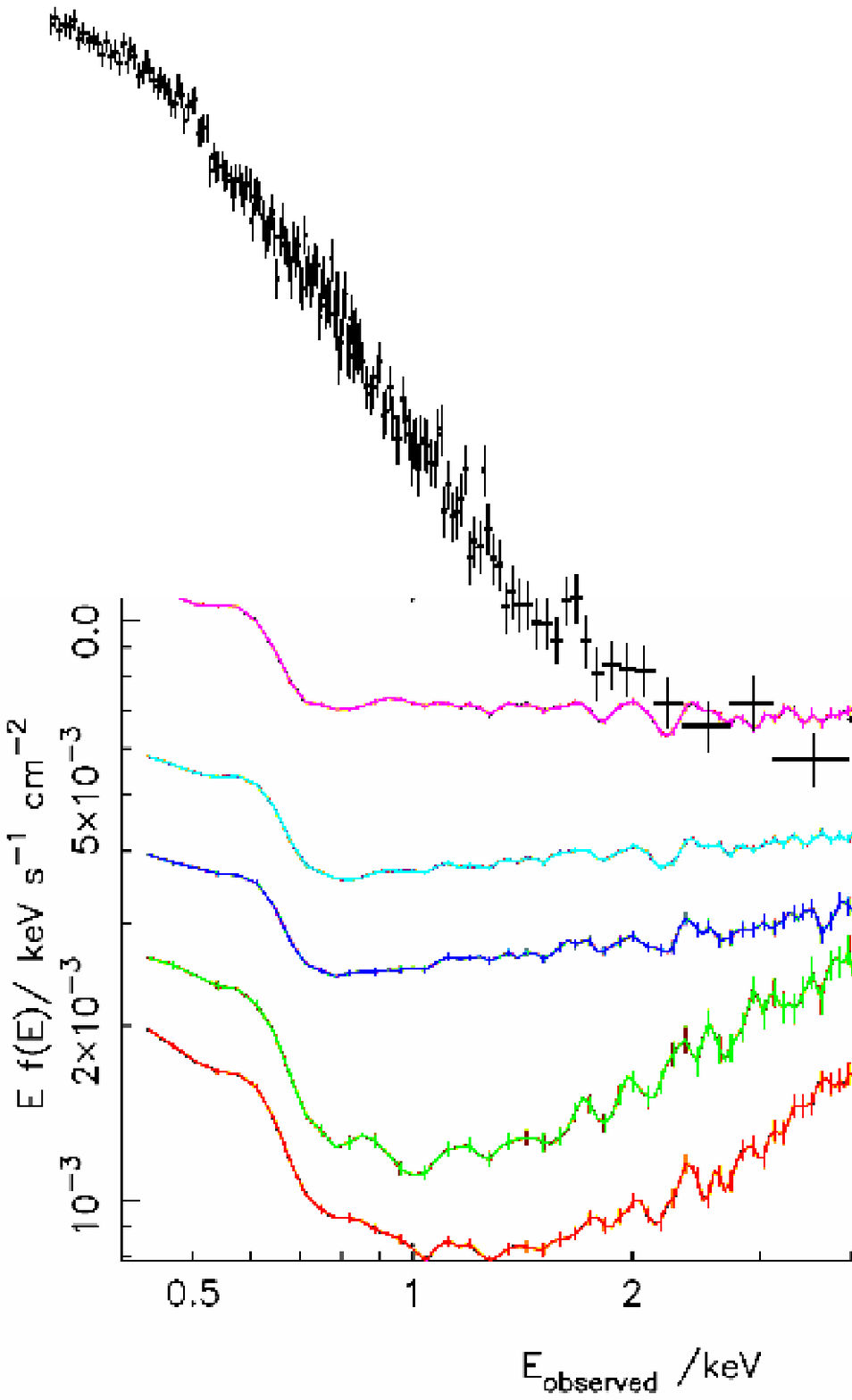}
\end{tabular}
\caption{As in Fig.~\ref{f:mkn766}, with Mkn766 as background but now
for objects showing somewhat different variability.
a) shows NGC~3516 (Turner et al 2008). Plainly there is some contribution
here from complex absorption.  b) shows 1H0707-495, which has a hardest
spectrum which is similar to that from Mkn766, but with an enourmous drop
above 7~keV (Boller et al 2002), and an extremely strong soft excess.
However, the softer
spectra at higher luminosity look more like the strange spectrum seen from
PDS456 (Fig.~\ref{f:mkn766}d), but also have 
clear features from iron L emission (Fabian et al 2009).
c) shows RE~J1034+396 one of  
the strongest soft excess known, which may be a real continuum component
(Middleton et al 2009).}
\label{f:agn_variety}
\end{figure*}

However, the X-ray spectrum is {\em not} simply made up of the power
law and its reflection. There is also a 'soft X-ray excess', clearly
seen in the spectra of most high mass accretion rate AGN
below 1~keV. Fig.~\ref{f:mkn766}a shows the changing shape of this
soft excess for different intensity sorted spectra of Mkn766.
This is rather smooth, so looks like a separate continuum 
component. However, fitting this for a large sample of AGN gives a 
typical 'temperature' of this component that is remarkably 
constant 
irrespective of mass of the black hole (Czerney et al 2003; Gierlinski
\& Done 2004b). This is very unlike the behaviour of a disc or any
component connected to a disc, requiring some (currently unknown)
'thermostat' to maintain this temperature. Another, more subtle,
problem is that 'normal' BHB do not show such a component in their
spectra, so these AGN spectra do not exactly correspond to a scaled up
version of the BHB high/very high states. However, there {\em is} such
a separate component in the most luminous state of the brightest BHB
GRS~1915+105 which may scale up to produce the soft excess in the most
extreme AGN (Middleton et al 2009; see Fig.~\ref{f:agn_variety}c).

More clues to the nature of the soft excess can come from its
variability. These objects are typically highly variable, and the
spectrum changes as a function of intensity in a very characteristic
way. At the highest X-ray luminosities, the 2-10~keV spectrum can
often be well described by a $\Gamma\sim 2.1$ power law, with resolved
iron emission line, and a soft excess which is a factor $\sim 2$ above
the extrapolated 2-10~keV power law at 0.5~keV.  Conversely, at the
lowest luminosities, the apparent 2-10 keV power law index is much
harder (and there are absorption systems from highly ionized iron),
and the soft excess above this extrapolated emission is much larger.
Fig.~\ref{f:mkn766}a shows this spectral variability for Mkn 766.

The entire spectrum at the lowest luminosity looks like moderately
ionized reflection (with the iron K alpha He and H like lines
absorption lines superimposed), but the lack of soft X-ray lines (as
well as the lack of a resolved iron emission line) means it would have
to be extremely strongly distorted by relativistic effects. The
classic extreme iron line source, MCG-6-30-15 has similar spectral
variability but with stronger 'warm absorption' features around 0.7-1~keV
and stronger
ionized iron absorption lines (see Fig.~\ref{f:mkn766}b).  In both
these sources, most of the spectral variability can be modeled if 
there is an extremely relativistically smeared reflection component which
remains constant, while the
$\Gamma=2.1$ power law varies, giving increasing dilution of the
reflected component at high fluxes (Fabian \& Vaughan 2003).

The obvious way for the reflected emission to remain constant is if it
is produced by far off material, but the extreme smearing requirement
conflicts with this. Instead, lightbending from a source very close to
the event horizon could give both the required central concentration
of the illumination pattern and apparent constancy of reflection if
the variability is dominated by changes in source position giving
changes in lightbending (Miniutti et al 2003; Fabian et al 2004;
Miniutti \& Fabian 2004). As the X-ray source gets closer to the black
hole, the X-ray emission is increasingly focused onto the inner disc,
and so the amount of direct emission seen drops.

However, there are some physical issues with this solution. Firstly,
unlike the iron line, the soft excess requires extreme smearing (small
inner radius, strongly centrally concentrated emissivity) in almost
all the objects in order to smooth out the strong soft X-ray line
emission predicted by reflection (Crummy et al 2006). 
Secondly, to produce the soft excess from reflection
requires that the reflecting material is moderately ionized over much
of the disc photosphere.  Yet the ionisation instability for material
in pressure balance means that only a very small fraction of the disc
photosphere can be in such a partially ionised state (see
section~\ref{s:instability}).  Either the disc is not in hydrostatic
equilibrium (held up by magnetic fields?) or the soft excess is not
formed from reflection (Done \& Nayakshin 2007).

Instead, it is possible that the soft excess is formed by partially
ionised material seen in absorption through optically thin material
rather than reflection from optically thick material. However, there
is still the issue of the lack of the expected partially ionised lines
- in absorption this time rather than from emission (see
Fig.~\ref{f:ionise}b). These could be smeared out in a similar way to
reflection if the wind is outflowing (Gierlinski \& Done 2004b;
Middleton et al 2007), but the outflow velocities required are
similarly extreme to the rotation velocities (Schurch \& Done 2008;
Schurch et al 2009). Even including scattering in a much more
sophisticated wind outflow model shows that 'reasonable' outflow
velocities are insufficient to blend the atomic features at low
energies into a pseudo-continuum, though this can explain the broad
iron line shape without requiring extreme relativistic smearing (Sim
et al 2010).

Potentially a more plausible geometry is if the absorber is clumpy,
and only partially covers the source, then there are multiple lines of
sight through different columns. Any mostly neutral material gives
curvature underneath the iron line, making an alternative to extreme
reflection for the origin of the red wing (L. Miller et al 2007; 2008;
Turner et al 2008). Such neutral material will produce an iron
fluorescence line, but this line can also be absorbed, so current
observations cannot yet distinguish between these two models (Yaqoob
et al 2009). This much more messy geometry perhaps gives more
potential to explain the larger range of complex variability seen in
some of the other high mass accretion rate AGN, as shown in
Fig.~\ref{f:agn_variety}a and b.

It is clearly important to find out which one of these geometries we
are looking at. Either we somehow have a clean line of sight down to
the very innermost regions of the disc despite it being an intense UV
source which should be powering a strong wind, or we are looking
through a material in strong, clumpy, wind, which has important
implications for AGN feedback models, giving another way to strongly
suppress nuclear star formation. These questions are currently an area
of intense controversy and active research.

\subsection{Observations of 
Reflected Emission and Relativistic smearing in BHB}
\label{s:bhobs}

Reflection is also seen in BHB, and here the controversy over its
interpretation occurs at both low and high mass accretion rates.  

The amount of reflection is determined by the solid angle subtended by
the disc to the hard X-ray source i.e. the fraction of the sky that is
covered by optically thick material as viewed from the hard X-ray
emission region. The truncated disc models for low mass accretion
rates predict that this should increase as the disc moves
in towards the last stable orbit, identified with the source making a
transition from the low/hard to high/soft state, while the decrease in inner
disc radius means that this should also be more strongly
smeared by relativistic effects. With RXTE data, the solid angle of
reflection increases as expected, but the poor spectral resolution
means that it is very difficult to constrain the relativistic
smearing. Nonetheless, these do appear more smeared in the RXTE data
(Gilfanov, Churazov \& Revnivtsev 1999; Zdziarski, Lubinski \& Smith
1999; Ibragimov et al 2005, Gilfanov 2010).

These results were derived assuming neutral reflection, whereas the
reflected spectrum is plainly ionised in the high/soft states
(e.g. Gierlinski et al 1999). While some of this ionization can be
from photo-ionization by the illuminating flux, at least part of it
must be due to the high disc temperature (i.e. collisional ionization:
Ross \& Fabian 2007). Thus the high/soft and very high state require
fitting with complex ionised reflection models in order to disentangle
the relativistic smearing and solid angle from the ionization state.
Nonetheless, attempts at this using simplisitic models of ionized
reflection ({\sc pexriv}) with the RXTE data gave fairly consistent
answers.  The high/soft data seemed to show that the solid angle
subtended by the disc is of order unity, and the (poorly constrained)
smearing gives $r_{in}\approx 6$ for emissivity fixed at $\beta=3$, as
expected from the potential geometries sketched in
Fig.~\ref{f:cygx1_soft}b (Gierlinski et al 1999, Zycki, Done \& Smith
1998).  By contrast, the very high state geometry discussed in
section~\ref{s:hybrid} required that the inner disc is covered by an
optically thick corona, predicting a smaller amount of reflection and
smearing, again consistent with the RXTE observations (Done \& Kubota
2006).

However, the first moderate and good spectral resolution results
appeared to conflict with the neat picture described above. The extent
of this is best seen in Miller et al (2009), who compiled some
XMM-Newton spectra of BHB (plus a few datasets from other satellites)
and fit with the best current relativistically smeared, ionised
reflection models (together with a power law and disc spectrum). Their
Table 3 shows that all the very high state spectra ($\Gamma>2.4$),
require a large solid angle of reflection from the very inner regions
of the disc, at odds with the geometry proposed from the continuum
shape where an optically thick corona completely covers the cool
material (Done \& Kubota 2006).

There are even worse conflicts with the models for the low/hard
state. The truncated disc/hot inner flow models clearly predict that
the amount of relativistic smearing should be lower than in the
high/soft state. However, XMM- Newton data from low/hard state
observations also show a line which is so broad that the disc is
required to extend down to the last stable orbit of a high spin black
hole. The most famous of these is GX339-4 (Miller et al 2006b; Reis et
al 2008), though this one is probably an artifact of instrumental
distortion of the data due to pileup (Fig.~\ref{f:lowhard}a, Done \&
Diaz-Trigo 2010). However, there are other data which also show a line
in the low/hard state which is so broad that the disc is required to
extend down to the last stable orbit of a rapidly spinning black hole
(SAX J1711.6-3808: Miller et al 2009; GRO~J1655-40 and XTE~J1650-500:
Reis et al 2009, see also Reis et al 2010). While these are not so
compelling as the (piled up) data from GX339-4, they still rule out
the truncated disc/hot inner flow models for the low/hard state if the
extreme broad line is the only interpretation of the spectral shape.

\section{The nature of the low/hard state in BHB and AGN}

This conflict motivates us to look again at the low/hard state in
particular, especially as there are other observations
which also challenge the hot inner flow/truncated
disc geometry. Again, this is currently  an area of
intense controversy and active research. 

\subsection{Intrinsic disc emission close to the transition}

The high/soft state disc dominated spectra trace out $L \propto T^4$
giving strong evidence for a constant size scale inner radius (see
section \ref{s:lt}). After the transition to the low/hard state there
is still a (weak) soft X-ray component which can be seen in CCD
spectra (but not in RXTE due to its low energy bandpass limit of
3~keV). This has temperature and luminosity which is consistent with
the same radius as seen in the high/soft state data, implying that the
disc does not truncate (Rykoff et al 2007, Reis et al 2010).

\begin{figure*}
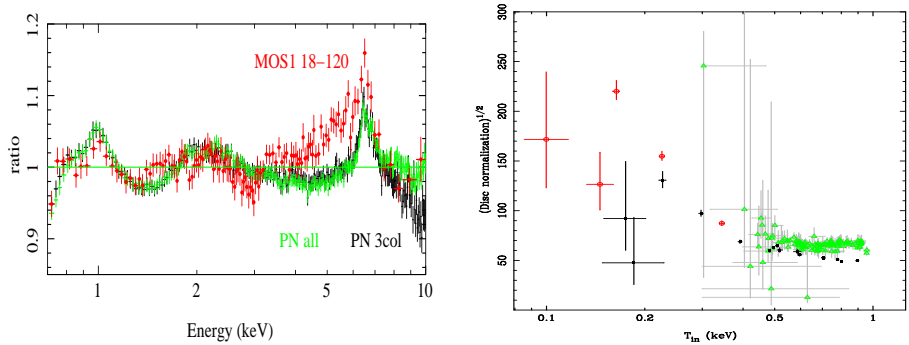
 
\begin{tabular}{cc} 
\includegraphics[width=0.45\textwidth, height=4.5cm]{pnall_pn3col_mos1.ps} 
&
\includegraphics[width=0.45\textwidth, height=4.5cm]{1817_temp_radius.ps}
\end{tabular} 
\caption{a) Residuals to a continuum model fit to the low/hard state
of GX339-4.  The red points show the MOS data on XMM-Newton, with an
obvious, extremely broad red wing to the iron line (Miller et al
2006b).  However, these data suffer from pileup, and the simultaneous
PN instrument on XMM-Newton which is not piled up, show a much
narrower profile (black and green data: Done \& Diaz-Trigo 2010)
b) Temperature versus scaled
inferred inner disc radius for SWIFT (black) and RXTE (green) data
from XTE J1817-330 fit to a disc plus Comptonization model. The source
is in the high/soft state for temperatures above 0.4~keV, as shown by
the constant radius.  The two lowest temperature points at $\sim
0.2$~keV are low/hard states, while the ones in the 0.2--0.4~keV range
are taken during the transition (intermediate state). The 3~keV lower
limit to the RXTE bandpass means that it cannot follow the disc as the
source goes into the transition, but the SWIFT data show that the disc
radius starts to increase.  A simple interpretation of the low/hard
state data are that the disc returns to the last stable orbit, but the
red points show how this can change by including irradiation, and the
radius can be further increased to 250-300 in these units by going
from a stress-free inner boundary to stress across the truncated disc
edge. The disc radius cannot be unambiguously determined from the
low/hard state data (Gierlinski et al (2008). }
\label{f:lowhard} 
\end{figure*}

Fig.~\ref{f:lowhard}b illustrates this with both RXTE (green) and SWIFT (black) data
from the transient XTE~J1817-330 (Rykoff et al 2007, Gierlinski, Done \& Page 2008). 
The disc goes below
the RXTE bandpass just as the source makes a transition from the high/soft state, 
but its evolution can be followed by the lower energy data from SWIFT. 
Plainly the SWIFT data show that the disc {\em does} recede during (i.e triggering!) the
transition, but then in the low/hard state proper, it apparently bounces back to give the same
radius as in the high/soft state. However, just after the transition,
the disc is not too far recessed, so it can be strongly illuminated by
the energetically dominant hard X-ray source. This changes the
temperature/luminosity relation from that expected from just
gravitational energy release, as shown by the red points in Fig.~\ref{f:lowhard}.
Additionally, the difference in inner
boundary condition (going from the stress-free last stable orbit to a
continuous stress across the truncated inner disc radius) means that
the same temperature/luminosity relation implies the disc is bigger,
which would move the two low/hard state points up to 250-300 in these units! 
This is even before taking into account that some of the disc
photons are lost to our line of sight through Compton scattering if
the disc underlies some of the hot inner flow. Putting these photons
back into the disc gives higher luminosity/larger radii. (Makishima et
al 2008).

These data show the difficulty in unambiguously reconstructing the inner radius 
of the disc when the disc component does not dominate the spectrum. They can be 
consistent with the disc down at the last stable orbit (Rykoff et al 2007; Reis 
et al 2010), but they can equally well be consistent with a truncated disc
(Gierlinski et al 2008). 
However, the data during the transition are fairly clear that the  
disc starts to recede, 
so it seems most likely to me that that this 
continues into the low/hard state.

\subsection{Intrinsic disc emission at very low luminosities}
\label{s:rep}

The transition is going to be complicated. The disc surely does not truncate in 
a smooth way, so there can be turbulent clumps, as well as issues with the 
overlap region suppressing the observed disc while also giving rise to strong 
illumination as discussed above. Instead, a much cleaner picture should emerge 
instead from a dimmer low/hard state if the disc truncates, as here it should be 
far from the X-ray source, so irradiation and overlap effects should be 
negligible. Yet there is still a weak soft X-ray component, with temperature and 
luminosity such that the emitting area implied is very small, of order the size 
scale of the last stable orbit. This has now been seen in several different CCD 
observations of low $L/L_{Edd}$ sources so is clearly a robust result (Reis, Miller
\& Fabian 2009; Chiang et al 2009; Wilkinson \& Uttley 2009)

However, putting the disc down to the last stable orbit is not a solution to 
these data. It is then  co-spatial with the hard X-ray emission, as this must be 
produced on small size scales. This then runs into difficulties with 
reprocessing. If the hard X-ray corona overlies the optically thick disc and 
emits isotropically then half of the hard X-ray emission illuminates the disc. 
Some fraction, $a$, (the albedo) of this is reflected, but the remainder is 
thermalized in the disc and adds to its disc luminosity. The minimum disc 
emission is where there is no intrinsic gravitational energy release in the 
disc, only this reprocessed flux. This is $L_{rep}=(1-a) L_h/2 \sim L_h/3$ as 
the reflection albedo cannot be high for hard spectral illumination as high 
energy X-rays cannot be reflected elastically. Instead they deposit their energy 
in the disc via Compton downscattering. Yet we see $L_{soft} \sim L_h/20$. Thus 
the geometry must be wrong! Either the hard X-ray source is not isotropic, 
perhaps beamed away from the disc as part of the jet emission, or the material 
is a small ring rather than a disc so that its solid angle is much less than 
$2\pi$ for a full disc (Chiang et al 2009; Done \& Diaz-Trigo 2010). 

However, for one source, XTE J1118+480, the galactic column density is
so low that there are simultaneous UV and even EUVE constraints on the
spectrum (Esin et al 2001). These show that this soft
X-ray component co-exists with a much more luminous, cooler UV/EUV
component. If the soft X-rays are the disc, what is the UV/EUV component?
Alternatively, since the UV/EUV component looks like a 
truncated disc, what is the soft X-ray component? 
It must come from a much smaller emission
area than the UV/EUVE emission, and one potential origin is the inner edge
(rather than top and bottom face) of the truncated disc 
(Chiang et al 2009). 
The truncation region is probably highly turbulent, so there can be 
intrinsic variability produced by clumps forming and shreding, as well
as them reprocessing the hard X-ray irradiation. This may also explain the 
variability seen in this component (Wilkinson \& Uttley 2009). 

Thus in my opinion, none of the current observations require that there is a 
disc down to the last stable orbit in the low/hard state. More fundamentally, it 
is very difficult to make such a model not conflict with other observations. 
Reprocessing limits on the hardness of the spectrum requires that the X-ray 
source is either patchy or beamed away from the disc (Stern et al 1995; 
Beloborodov 1999). A patchy corona would give a reflection fraction close to 
unity, which is not observed even considering complex ionization of the reflected
emission
(Barrio et al 2003; Malzac et al 2005). Beaming 
naturally associates the X-ray source with the jet, but is unlikely to be able 
to simultaneously explain the extreme broad line parameters derived for some 
low/hard state sources since the illumination pattern becomes much less 
centrally concentrated by the beaming. I suspect that more complex continuum 
modelling may make these lines less extreme, but this then removes a challenge 
to the truncated disc as well! And unlike the beaming models, the truncated 
disc/hot inner flow can additionally give a
mechanism for the major state tranistions, and the variability.

\section{Conclusions}

The intrinsic radiation processes of blackbody radiation and
Comptonization go a long way to explaining the underlying
optical-to-X-ray continuum seen in both BHB and AGN. These, together
with the atomic processes of absorption and reflection, and 
relativistic effects in strong gravity give us a
toolkit with which to understand and interpret the spectra of the
black hole accretion flows. This is currently an area of intense 
and exciting research, to try to understand accretion in strong gravity. 
If you got this far, congratulations, and come and join us: we get to
play around black holes! 

\section{Acknowledgements}

I would like to thank the organizers of the IAC winter school for
inviting me to give this series of lectures, finally giving me the
motivation to write these things down. But I only know these things
because of the many people who have given me their physical insight on
radiation processes, especially Andy Fabian and Gabriele
Ghisellini. I also thank ISAS and RIKEN for visits during which I 
developed some of these lectures. 
It also could not have been written without the 8 hours
spent at Tenerife South Airport where their lack of baggage check
facility put paid to my plan to spend the day at a nearby surfing
beach!

\end{document}